\documentclass[12pt,preprint]{aastex}
\usepackage{flushrt,float}

\slugcomment{Submitted to the Astrophysical Journal - Revised Version}

\shortauthors{Chieffi, Limongi} \shorttitle{Core Collapse Supernovae}

\begin{document}

\title{Evolution, Explosion and Nucleosynthesis of Core Collapse Supernovae}

\author{Marco Limongi\altaffilmark{1} and Alessandro Chieffi\altaffilmark{2}}

\affil{1. Istituto Nazionale di Astrofisica - Osservatorio Astronomico di Roma, 
Via Frascati 33, I-00040, Monteporzio Catone, Italy; marco@mporzio.astro.it}

\affil{2. Istituto di Astrofisica Spaziale e Fisica Cosmica (CNR), Via Fosso del 
Cavaliere, I-00133, Roma, Italy; achieffi@rm.iasf.cnr.it}

\begin{abstract}

We present a new set of presupernova evolutions and explosive yields of massive 
stars of initial solar composition (Y=0.285, Z=0.02) in the mass range 13-35 
$\rm M_\odot$. All the models have been computed with the latest version (4.97) 
of the FRANEC code that now includes a nuclear network extending from neutrons 
to $\rm ^{98}Mo$. The explosive nucleosynthesis has been computed twice: a first 
one with an hydro code and a second one following the simpler radiation 
dominated shock approximation (RDA). The main results concerning the models and 
the associated explosions are: a) the inclusion of the latest available input 
physics does not alter significantly the main properties of the presupernova 
models with respect to our previous ones; b) the differences between the old and 
new explosive yields computed with the RDA remain confined within a factor of 
two for most of the nuclei; c) the differences between the elemental RDA yields 
and those computed with an hydro code do not exceed 0.1 dex with the exceptions 
of the elements produced mainly by the complete explosive Si burning - in this 
case the variations are anyway less than a factor of three; d) the relation 
between the final kinetic energy and the $\rm ^{56}Ni$ ejected weakens as the 
mass of the progenitor star increases; e) the yields corresponding to the 
ejection of a given amount of $\rm ^{56}Ni$ are in fair agreement with those 
obtained by a more energetic explosion in which the mass cut is chosen in order 
to give the same $\rm ^{56}Ni$. The production factors (integrated over a 
Salpeter mass function between 13 and 35 $\rm M_\odot$) of the majority of the 
isotopes show an almost scaled solar distribution relative to Oxygen. The few 
exceptions are discussed in details. We also find that the present integrated 
yields allow a fraction of the order of 10-20\% of SNIa if their latest 
available yields are adopted. \end{abstract}



\section{Introduction}

Some years ago we started a big project whose main objective was the study of 
the evolutionary properties of the massive stars and of their explosive yields. 
In the first paper of this series \cite[hereinafter CLS98]{cls98} we mainly 
discussed the FRANEC code and the presupernova evolution of a $\rm 25 M_\odot$ 
star of solar metallicity. The nuclear network adopted in that paper extended up 
to $\rm ^{67}Zn$ and included 12 nuclei in the H burning, 25 in He burning and 
149 in the more advanced phases. In the second paper \cite[hereinafter 
LSC00]{lsc00} we presented the evolutionary properties of a set of solar 
metallicity stars (namely $\rm 13, 15, 20, 25~M_\odot$) with a network still 
extending up to Zn but with 41 nuclei in the H burning, 88 in the He burning and 
179 in the more advanced burning phases. In the same paper we also presented the 
explosive yields (in the frame of the radiation dominated shock approximation) 
produced by stellar models of different initial metallicities, namely Z=0.02, 
Z=0.001 and Z=0. In the following papers (Limongi \& Chieffi 2002; Chieffi \& 
Limongi 2002a, hereinafter CL02a) we discussed in detail the yields produced by 
the first stellar generation and used them to interpret the observed surface 
abundances of the available ultra metal poor stars. In \cite{cl02b} we addressed 
the production of the radioactive nuclei by solar metallicity massive stars.

Here we present a new generation of stellar models and associated explosions. 
The present pre explosive models differ from the previous ones in several 
respects: first of all the input physics has been updated whenever possible, 
second the nuclear network has been extended up to $\rm ^{98}Mo$, so that we now 
follow 44 nuclei in H burning, 149 in He burning and 267 in the more advanced 
phases. Further, distinctive differences between the present and the previous 
sets of models concern a) the adoption in the current models of the Ledoux 
criterium to determine the stability of the H convective shell that forms 
towards the end of the central H burning phase and b) the adopted rate for the 
$\rm^{12}C(\alpha,\gamma)^{16}O$ process. Though the effective rate for this 
process still shows an embarrassing large uncertainty, at variance with our 
previous works where we chose to adopt the rate quoted by \cite{cf85}, (but see 
Imbriani et al., 2001 and CL02a), in the present one we decided to adopt the 
best current estimate \citep{kunzetal02} at its face value without any attempt 
to rescale it in any way.

As for the computation of the explosive burnings, a big effort has been done to 
build up a hydro code capable of following the passage of a shock wave through 
the mantle of a star. Hence we can now compute the explosive nucleosynthesis by 
means of either the hydro code and the radiation dominated shock approximation: 
a comparison between the yields obtained with the two different approximations 
is discussed in detail. We have also explored at some extent the dependence of 
the explosive yields on the final kinetic energy of the ejected mantle.

A further improvement has been the extension of the network up to the N=50 
closure shell, which means the possibility to explore the production of the so 
called "weak component" of the {\em s}-processes. Let us briefly remind that the 
solar system distribution above A=60 is the result of the combined effects of 
{\em s}-, {\em r}- and {\em p}- processes, i.e. of slow and rapid neutron 
captures (with respect to the decay times of the unstable nuclei near stability) 
plus a contribution from the p- captures. The fit to the {\em s}-only (and quasi 
{\em s}-only) nuclei, i.e. those produced only by slow neutron captures, showed 
\citep{wnc76,wn78,kap82} that at least two different neutron fluxes are 
necessary to reproduce their observed abundances. A "strong" pulsed neutron 
exposure (called "main component"), usually associated to the thermal pulses 
occurring in the intermediate mass stars, would account for the {\em s}-nuclei 
above A=90 and a weaker single exposure (called "weak component"), associated 
with the massive stars, would be responsible for the production of the {\em s}-
nuclei between A=60 and A=90. The nucleosynthesis of the {\em s}-nuclei in the 
massive stars has been addressed up to now in a number of papers, namely 
\cite{csa74,lhti77,at85,bg85,paa87,laa89,phn90,rai91a,rai91b,bep92,rai93,tem00,hww01,rau02}.
In spite of this not negligible number, almost all of these papers 
considered only the contribution of the central He burning to the final yields, 
neglecting completely the possible contribution (positive and/or negative) of 
the further evolutionary phases {\em including the explosive nucleosynthesis}. 
Only two papers \citep{rai91b,rai93} added the contribution of the C-burning 
shell up to the end of the 90's. Just in two very recent papers the yields of 
the nuclei between A=60 and A=90 have been computed by considering the whole 
evolutionary history of the massive stars, including the explosive burnings, 
namely \cite{hww01} and \cite{rau02}. Our work proceeds on the same guideline, 
in the sense that all the nuclei of interest have been fully included in both 
the evolutionary and explosive codes, but with the difference that our 
computations extend to a wider mass grid, i.e. between 13 and 35 $\rm M_\odot$.

The paper is organized as follows: the latest version of the FRANEC code is 
described in section 2 while the hydrostatic evolution of the models is 
presented in section 3. The explosive nucleosynthesis computed by means of 
either the hydro code and the radiation dominated shock approximation is 
discussed in section 4. Section 5 presents a wide discussion of the present 
results. A final conclusion follows.

\section{Stellar evolution code and input physics}

The presupernova evolutions  presented in this  paper have been computed by 
means  of the latest version (4.97) of the FRANEC package. The most important 
changes with  respect to the previous release (CLS98 and LSC00) are  the 
following.

The nuclear network has been extended up to Molybdenum and now it includes 40 
isotopes (from neutrons to $\rm  ^{30}Si$) in hydrogen burning, 149 isotopes 
(from neutrons to $\rm ^{98}Mo$) in helium burning and 267 isotopes (from 
neutrons to $\rm ^{98}Mo$) in all the more advanced burning phases. In total 282 
isotopes (Table 1) and about 3000 reaction rates were explicitly included in the 
various nuclear burning stages. As in the previous versions, the nuclear network 
is fully coupled to the equations describing the physical  structure of the star 
so that both the physical and chemical evolution due to the nuclear reactions 
are solved simultaneously. This means that the full network is used to compute 
the nuclear energy generation rate for each meshpoint and at each timestep. The 
system of differential equations is now solved by means of the Yale sparse 
matrix package that provides a speed up factor of $\sim 2$  compared to  the LU 
decomposition (and subsequent backward  and forward substitution) previously 
adopted. We have also removed the NSE approximation for temperatures  larger 
than $\rm 4 \cdot 10^{9}~K$  so that the full network is now used until the 
beginning of the core collapse of the star.

All the nuclear reaction rates are taken from the latest version of the {\em 
REACLIB} \citep{RT2000} available also at T. Rauscher's home page
\footnote{$\it http://quasar.physik.unibas.ch/\sim tommy$}
with some exceptions described below. {\em REACLIB} is a nuclear reaction rate 
library that contains fits to both experimental and theoretical rates. The main 
difference between this and the previous version of {\em REACLIB} (adopted in 
CLS98 and LSC00) consists in an upgrade of the theoretical fits, as presented by 
\cite{RT2000}. The experimental rates were not updated. Whenever possible we 
always chose the experimental rates.

The nuclear cross sections for the (n,$\gamma$) reactions are derived from the 
database of experimental values provided by \cite{BK2000}. These data cover the 
energy (temperature) range between 5 and 100 keV ($5.80\cdot10^{7}$ -
$1.16\cdot10^{9}$ K). Above 100 keV we use the theoretical rates computed by 
\cite{RT2000}, but rescaled to match the experimental values at 100 keV. The 
reverse rates, i.e. the ($\gamma$,n) ones, are those provided by Rausher, 
rescaled by the same multiplier factors used for the forward reactions.

The $\rm ^{12}C(\alpha,\gamma)^{16}O$ and $\rm ^{22}Ne(\alpha,n)^{25}Mg$ rates
adopted here are the ones provided by \cite{kunzetal02} (the adopted rate) and
by \cite{jaegeretal01} (the recommended rate) respectively.

The $\rm ^{18}O(\alpha,\gamma)^{22}Ne$ and $\rm ^{22}Ne(\alpha,\gamma)^{26}Mg$
rates are the recommended values given by \cite{kapetal94} for $\rm T9\leq0.6$.
Above this temperature we match the \cite{ca88} analytical relations to the
experimental data.

The weak interaction rates are taken by \cite{odaetal94} for A=1 and $\rm 17\leq 
A\leq 39$; \cite{FFN82,FFN85} for $\rm 40\leq A\leq 45$ (i.e. the same adopted 
in CLS98 and in LSC00); \cite{lp00} for $\rm 46\leq A\leq 62$; \cite{ty87} for 
$\rm A\geq 63$; {\em REACLIB} for all the nuclei not included in the above 
mentioned collections.

The initial distribution of the heavy elements has been assumed to be scaled 
solar and the relative abundances of the nuclear species above He are derived 
from \cite{ag89}. As usual the initial He abundance has been determined by 
fitting the present properties of the Sun \citep{SCL97}.

The equation of state adopted in CLS98 and LSC00  for temperatures below $\rm
10^{6}~K$ has been replaced by  the latest EOS and EOSPLUS tables given by
\cite{rsi96} and \cite{ro01}. Above $\rm T=10^{6}~K$  we are still using the
same  EOS described in LSC00.

The opacity tables are the same used in CLS98 and in LSC00: the radiative 
opacity coefficients are derived from \cite{Kurucz91} for $T\leq 10^{4}~{\rm 
K}$, from \cite{IRW92} (OPAL) for $10^{4}~{\rm K}<T\leq10^{8}~{\rm K}$, and from 
the Los Alamos Opacity Library (LAOL) \citep{Huebneretal77} for $10^{8}~{\rm 
K}<T\leq10^{10}~{\rm K}$. The opacity coefficients due to the thermal 
conductivity are derived from \cite{Itohetal83}.

As in CLS98 and LSC00, the  borders of the convective zones are  defined by 
means of the Schwarzschild  criterion except for the H convective shell 
that forms towards the end of the central H-burning, where the Ledoux criterion 
has been adopted \citep{ietal01}. According to this stability criterion, the H 
"convective" shell remains essentially stable against the growth of convective 
motions so that no mixing occurs in this region. This choice is based on some 
observational constraints mainly related to the relative number of red and blue 
supergiants \citep{LM95}. No mass loss has been taken into account.

As a last comment let us emphasize that the present computations are much more 
numerically accurate than the CLS98 and LSC00 ones since they have been obtained 
with a higher degree of both spatial and temporal resolution: the typical number 
of meshpoints is increased from 600 to 1500 while the number of timesteps in 
the advanced burning phases is roughly doubled (from 10000 to 20000).

\section{The Hydrostatic Evolution}

\begin{figure} 
\epsscale{1.0} 
\plotone{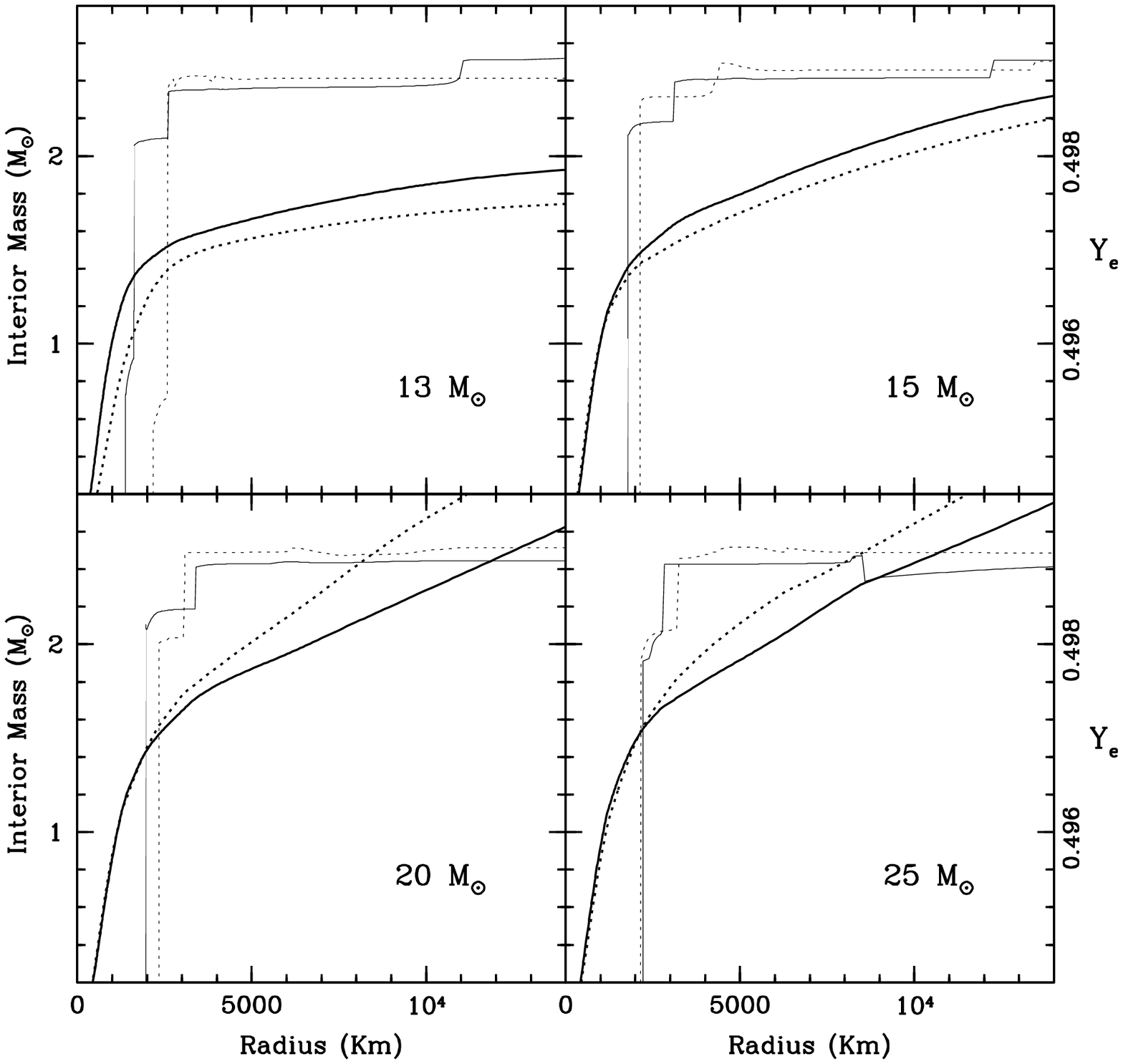} 
\caption{Presupernova Mass coordinate (thick lines) and electron mole number 
profiles ($Y_{\rm e}$) (thin lines) within the inner 14000 Km for the four 
masses in common between the present paper ({\em solid}) and LSC00 ({\em 
dotted}). \label{yemr}} 
\end{figure}

We computed the evolution of  6 models, namely 13, 15,  20, 25, 30 and $\rm 
35~M_\odot$, covering most of the mass interval that is limited at the lower end 
by the stars that ignite C in a partially degenerate environment and at the high 
one by the dominion of the Wolf Rayet stars. The initial chemical composition 
for all the  models is Y=0.285 and  Z=0.02. We have already largely  discussed 
the evolutionary properties of the various models from the pre main sequence 
phase  up to the final core collapse in earlier papers of this series (i.e. 
CLS98 and LSC00) and hence we will not repeat them here. For sake of 
completeness, let us mention that no breathing pulses develop during the last 
phase of central He burning because of  the very fine spatial and temporal 
resolution (about 3000 models and 2000 meshpoints just to follow the central  He 
burning phase). The key properties of these evolutions are reported in Table 2 
and may be compared to the older ones reported in Table 2 of LSC00. Such a 
comparison shows that the new models do not differ significantly with respect to 
the older ones: the largest differences occur for the $\rm 20~M_\odot$, but 
this is  due  to  the quite unusual behavior of the old $\rm 20~M_\odot$. The 
two key quantities  that play a fundamental  role in  the determination of the 
final explosive yields, i.e. the electron mole number $Y_{\rm e}$  and the 
final mass-radius relation within the region where these explosive  burnings 
occur, are shown in Figure \ref{yemr} as a function of the radial coordinate: 
the old and new relations are shown as dotted and solid lines respectively while 
the thin and thick lines refer, respectively, to $Y_{\rm e}$ and to the mass 
coordinate M. This figure shows that $Y_{\rm e}$ is more or less systematically 
lower by $\rm \sim 10^{-4}~$ in the new models while the M-R relation is now 
slightly steeper for the two smaller masses but somewhat shallower for the two 
more massive ones. However, the total mass  exposed to the various explosive 
burnings \citep{vulc2000} did not change by more than $\sim 30\%$ at most. 

\section{The Explosion}

The present modeling of the core collapse supernovae in spherical symmetry did not 
yield to successful explosions yet (at least with the presently available 
microphysics). Even improvements in the neutrino transport \citep{mezz01} and the 
inclusion of general relativistic corrections \citep{lieben01} in the model 
calculations did not change this situation. For this reason the explosive 
nucleosynthesis calculations for core collapse supernovae are still based on 
explosions induced by injecting an arbitrary amount of energy in a (also arbitrary) 
mass location and then following the development of the blast wave with an hydro 
code. The amount of energy injected is tuned in order to obtain a prefixed amount 
of kinetic energy of the ejecta at the infinity (usually $\rm \sim 10^{51}~erg$). 
Two different techniques are adopted at present: the thermal bomb and the piston. 
In the first approach the extra energy is deposited instantaneously in the form of 
internal energy at the base of the mantle of the exploding star. In the second one, 
conversely, the extra energy is given in the form of kinetic energy by assuming the 
inner edge of the mantle to move like a piston in the gravitational field of the 
compact remnant and to follow a ballistic trajectory. A third technique that may be 
adopted to compute the final explosive yields is the radiation dominated shock 
approximation (Weaver \& Woosley, 1980; Arnett 1996; CL02a) that we adopted in our 
previous works. In the following we will firstly present the yields obtained by 
adopting this last technique and compare them to the analogous ones presented in 
LSC00. We then present our hydro code together to the properties of the explosions. 
A comparison between the explosive yields obtained with the hydro code and the ones 
computed in the radiation dominated shock approximation will close this section.

\subsection{The radiation dominated shock approximation (RDA)}

\begin{figure}
\epsscale{1.0}
\plotone{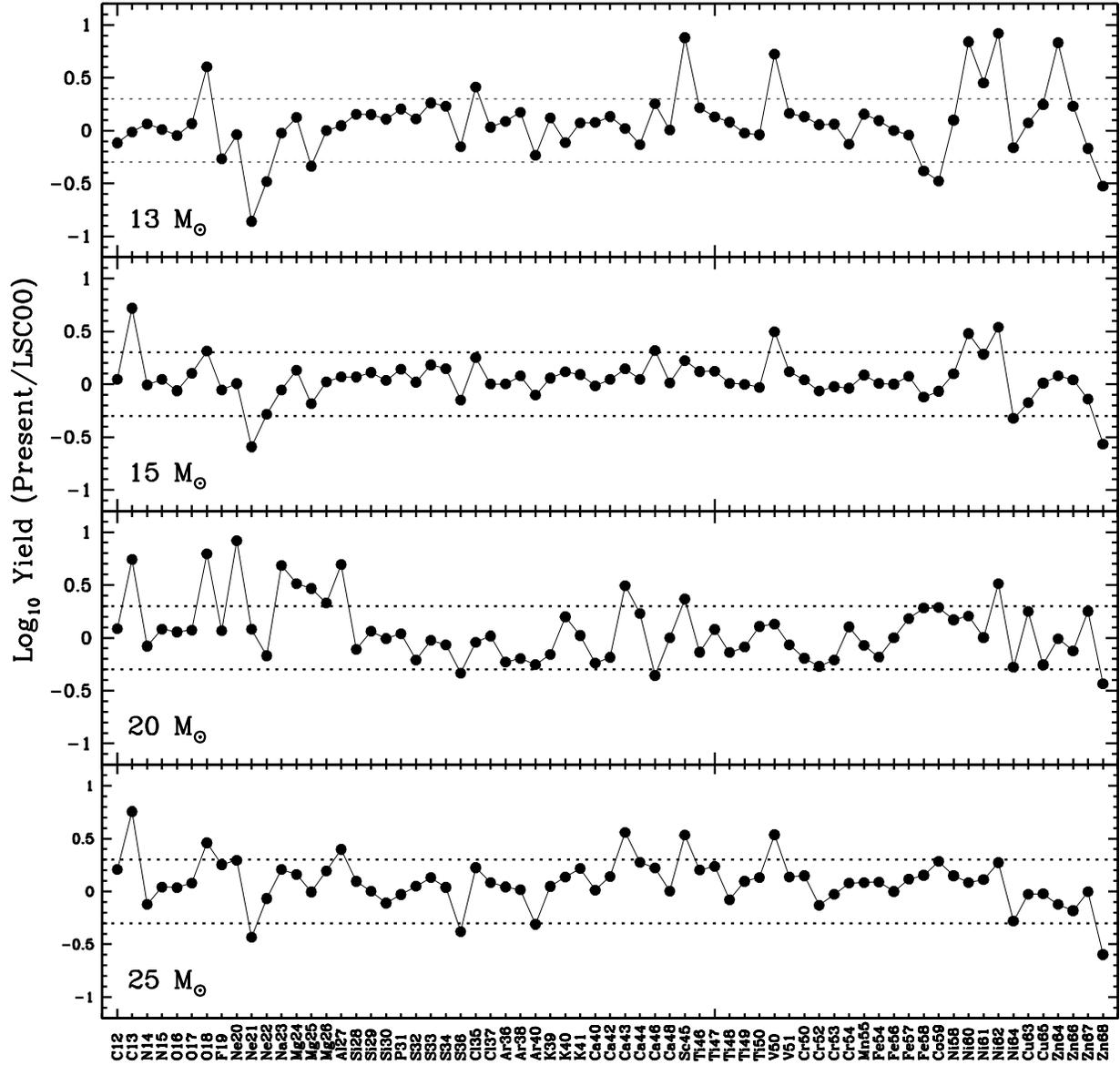}
\caption{Logarithmic  ratio  between the present and the LSC00 yields for the 
four masses in common. \label{oldnewrd}}

\end{figure}

A full set of explosive yields have been computed in the radiation dominated shock 
approximation by (arbitrarily) imposing a final kinetic energy of 1.2 foe ($\rm 
1~foe~=~ 1\times10^{51}~erg$), the ejection of $\rm 0.05~M_{\odot}$ of $\rm 
^{56}Ni$ and a time delay of 0.5 s between the beginning of the core collapse and 
the rejuvenation of the shock wave. Figure \ref{oldnewrd} shows the comparison, for 
the four masses in common, between the LSC00 and the present yields. Since the 
numerical technique adopted in both the older and newer simulations of the 
explosion is identical, the differences  are mainly due to the different 
presupernova structures and to the different nuclear reaction rates. This figure 
clearly shows that the differences remain confined within a factor of two for most 
of the isotopes, raising up to one order of magnitude for just few isotopes. Note 
that the largest differences between the new and the old computations do not show 
any specific clear trend with the mass and since the improvements to either the 
code and the input physics were applied all together, it is not easy to disentangle 
which is/are the specific changes that produced such differences. However, it is 
clear that the differences concerning the nuclei that are produced by the explosive 
burnings mainly reflect the differences in both the final Mass-Radius relation and 
the {\em Ye} profile. Viceversa the differences concerning the nuclei produced by 
the hydrostatic burnings will mainly reflect the changing in the reaction rates 
(the physical evolution of the stellar interiors did not change appreciably in the 
central H and He burning phases).

\subsection{The hydrodynamical explosion}

\subsubsection{The hydro code}

The propagation of a shock front through the mantle of a star is followed by 
solving the hydrodynamical equations in spherical symmetry and in lagrangean 
form:

\begin{equation}\label{eq1}
{{\partial v}\over{\partial t}}=-{{Gm}\over{r^2}}-4\pi r^2{{\partial P}\over{\partial m}}- 4 \pi {{\partial(r^2Q)}\over{\partial m}}
\end{equation}

\begin{equation}\label{eq2}
{{\partial r}\over{\partial m}}={{1}\over{4 \pi r^2 \rho}}
\end{equation}

\begin{equation}\label{eq3}
{{\partial e}\over{\partial t}}={{P}\over{\rho^2}}{{\partial \rho}\over{\partial t}}-4\pi r^2 Q{{\partial v}\over{\partial m}}
\end{equation}

supplemented by the boundary condition:

\begin{equation}\label{eq4}
{{\partial v_1}\over{\partial t}}=-{{Gm}\over{r^2}}
\end{equation}

that imposes the inner edge of the exploding mantle to move on a ballistic 
trajectory under the gravitational field of the compact remnant.

The technique adopted to solve the system of equations \ref{eq1}, \ref{eq2} and 
\ref{eq3}, is the same described by \cite{rm67} and \cite{mezz93}. Few 
additional zones are added to the computational domain in order to allow both 
the forward and the reverse shock waves to leave the structure. Once the shock 
enters one of these zones, they are removed from the computational domain, 
simulating in this way the escape of the shock wave out of the expanding mantle. 
The chemical evolution of the matter is computed by coupling the same nuclear 
network adopted in the hydrostatic calculations to the system of equations 
\ref{eq1}, \ref{eq2} and \ref{eq3}. The blast wave is started by imparting an 
initial velocity $v_{0}~$ at a mass coordinate of $\rm \simeq 1~M_\odot$, i.e. 
well inside the iron core, and $v_{0}$ is properly tuned in order to obtain a 
given final kinetic energy of the ejecta.

\subsubsection{Propagation of the shock, fall back and explosive nucleosynthesis}

\begin{figure}
\epsscale{0.8}
\plotone{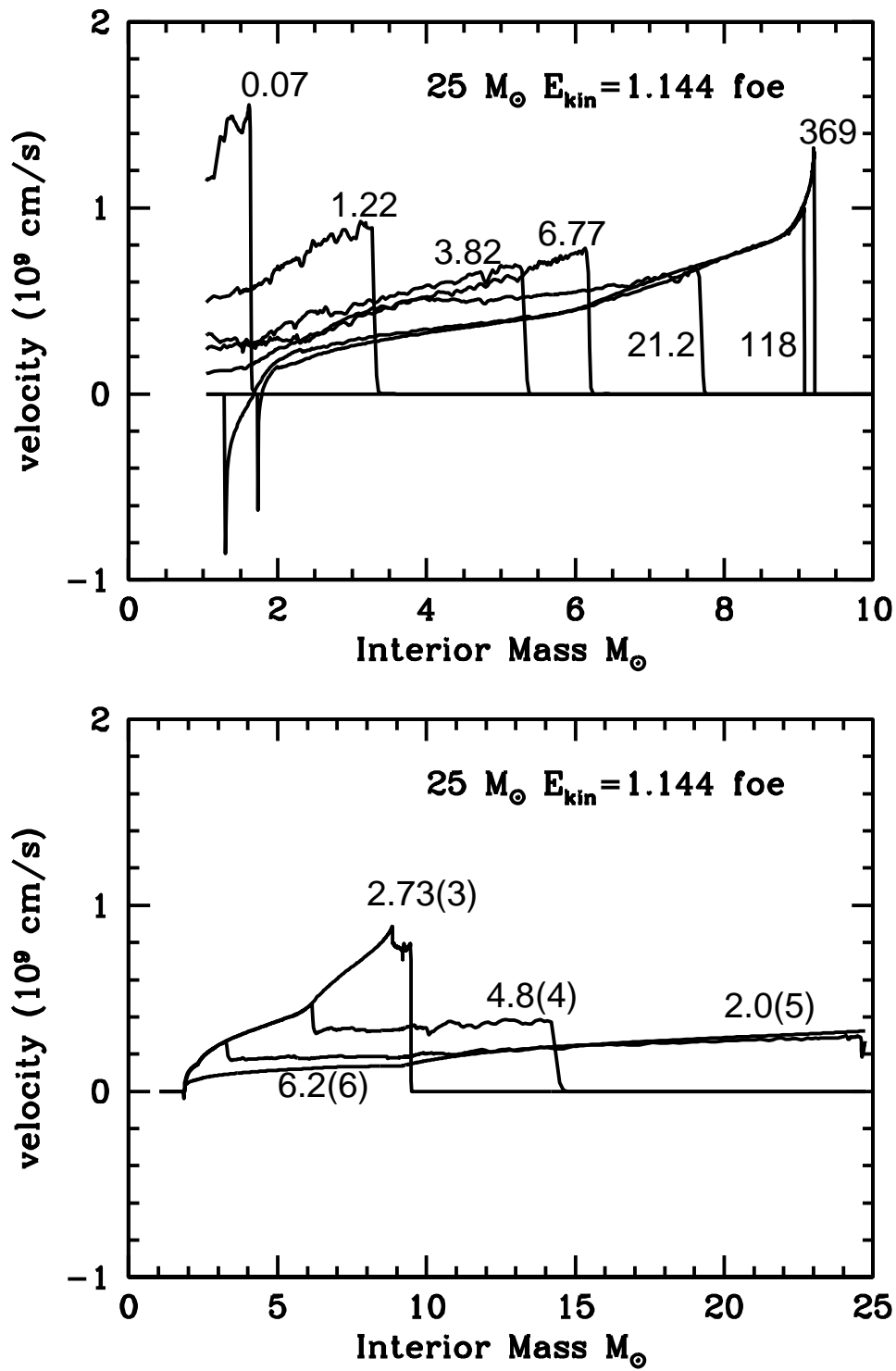}
\caption{Time history of the shock propagation in a $\rm 25~M_\odot$ model 
characterized by an initial velocity $v_{0}=1.555\cdot10^{9}~\rm cm/s$ and by a 
final kinetic energy at the infinity of $\rm 1.144\cdot10^{51}~erg$. Each curve 
is labeled by the time, in seconds, at which it refers. \label{shockprop}}
\end{figure}

The main properties of the explosion of the presupernova models presented in 
this paper can be summarized by discussing one specific model taken as a 
representative case. Figure \ref{shockprop} shows the time history of the shock 
propagation in a $\rm 25~M_\odot$ model characterized by an initial velocity 
$v_{0}=1.555\cdot10^{9}~\rm cm/s$ and by a final kinetic energy at the infinity 
of $\rm 1.144\cdot10^{51}~erg$. Once the shock forms, it propagates outward in 
mass increasing locally both the temperature and the density, triggering 
therefore the explosive nucleosynthesis. Behind the shock front both the 
pressure and the density (and hence the temperature) are fairly flat (occurrence 
largely exploited in the radiation dominated shock approximation) and 
progressively lower as the matter kicked by the shock wave expands and cools 
down. In $\rm \simeq 4~s$ the shock reaches the outer edge of the CO core; at 
this time the temperature in the shocked region has dropped down to $\rm 
10^{9}~K$ and the explosive nucleosynthesis almost vanishes. At $\rm t\simeq 
100~s$ the inner zones of the mantle begin to fall back onto the compact remnant 
while the shock is still within the He core. The fall back lasts $\rm 
\simeq1\cdot10^{5}~s$ though most of the mass that will eventually return to the 
remnant actually falls back in just $\rm \simeq 200~s$. The shock reaches the 
He/H interface at $\rm t\simeq 370~s$ and since this region corresponds to a 
very steep increase of $\rho r^{3}$, a reverse shock forms \citep{b90}. This 
reverse shock propagates inward in mass and decelerates somewhat the expanding 
matter that encounters in its way back: however, at least in our computations, 
this reverse shock does not affect significantly the amount of fall back. The 
main outgoing shock reaches the surface of the star at $\rm 
t\simeq2\cdot10^{5}~s$, while the reverse shock escapes from the inner edge of 
the mantle at $\rm t\simeq6\cdot10^{6}~s$. The further evolution of the mantle 
is characterized by an homologous expansion with velocity ranging between 1000 
(inner zones) and 3000 Km/s (outer zones).

In order to explore the dependence of the explosion on the initial velocity 
$v_{0}$ we have computed several hydro simulations that are summarized in Table 
3. The legend is as follows: the initial mass (column 1), a label to easily 
identify the model in the subsequent tables (column 2), the initial velocity 
$v_{\rm 0}$ (column 2), the times at which the shock reaches the CO core (column 
3), the fall back starts (column 4) and stops (column 5), the shock reaches the 
He/H interface (or equivalently the time at which the reverse shock forms) 
(column 6), the forward shock escapes from the surface (column 7) and the 
reverse shock escapes from the inner edge of the mantle (column 8), the final 
kinetic energy at the infinity (column 9), the mass of the compact remnant 
(column 10) and the amount of $\rm ^{56}Ni$ synthesized (column 11). The final 
explosive yields produced by most of these hydro simulations are reported in 
Tables 4-9 once the unstable isotopes have been decayed into their stable 
closest neighbor. The yields of all the $\gamma$-ray emitters (having a yield 
larger than $\rm 10^{-10}$ solar masses) included in the network are collected 
in Table 10.

An interesting result that comes out from these hydro simulations is that the 
correlation between the amount of $\rm ^{56}Ni$ ejected and the final kinetic 
energy of the ejecta weakens as the mass of the progenitor increases. Let us 
consider, for each mass, two limiting final kinetic energies $\rm E_{\rm kin}$: the 
first one, $\rm E^{\rm max}_{\rm kin}$, is the maximum kinetic energy that prevents 
the ejection of part of the iron core while the second one, $\rm E^{\rm min}_{\rm 
kin}$, is the minimum kinetic energy necessary to eject at least $\rm 10^{-
3}~M_\odot$ of $\rm ^{56}Ni$. The range of kinetic energies defined in this way is 
much larger for the lower mass models than for the more massive ones. For example 
the $\rm \Delta (E^{\rm max}_{\rm kin} - E^{\rm min}_{\rm kin}) \simeq 0.63~foe$ 
for the $\rm 13~M_\odot$ model while it reduces to $\rm \Delta (E^{\rm max}_{\rm 
kin} - E^{\rm min}_{\rm kin}) \simeq 0.2~foe$ for the $\rm 35~M_\odot$ model. 
Viceversa the maximum amount of the $\rm ^{56}Ni$ ejected remains roughly constant 
over the mass range 13-30 ${\rm M_\odot}$ and increases a while in the 35 ${\rm 
M_\odot}$ case. This means that explosions leading to the ejection of significantly 
different amounts of $\rm ^{56}Ni$ imply also a significative scatter in the final 
kinetic energy of the mantle if the mass is not very large, while they would imply 
a substantially {\it constant} final kinetic energy if the star is massive enough.

\begin{figure}
\epsscale{0.9}
\plotone{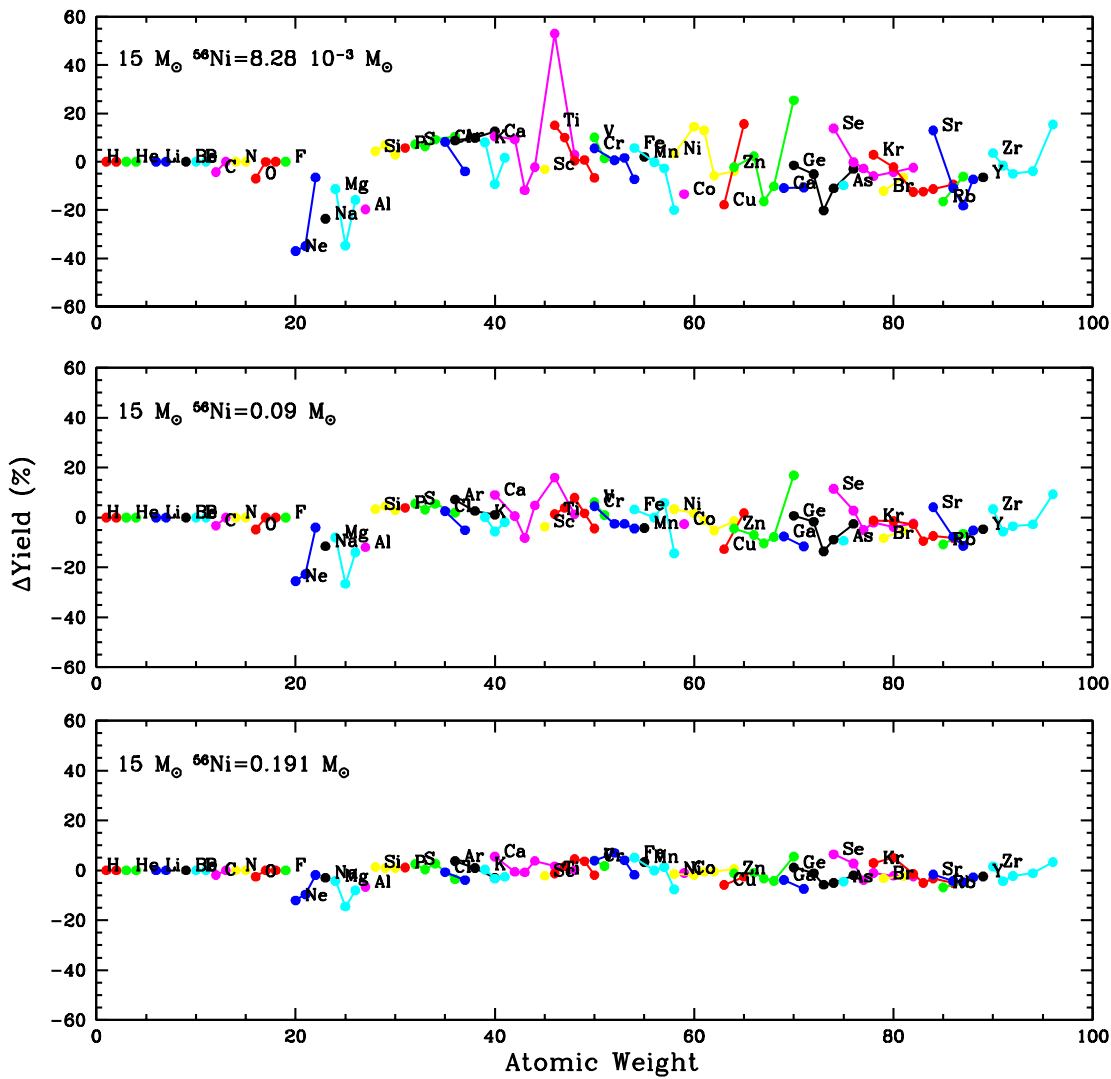}
\caption{Comparison between the yields of the $\rm 15~M_\odot$ star
computed for three specific explosions 
and the ones obtained for the run with the largest kinetic
energy but with the mass cut chosen to provide in each case the same 
$\rm ^{56}Ni$. Isotopes of the same elements are connected by a line.
\label{yield15comp}}
\end{figure}

\begin{figure}
\epsscale{0.9}
\plotone{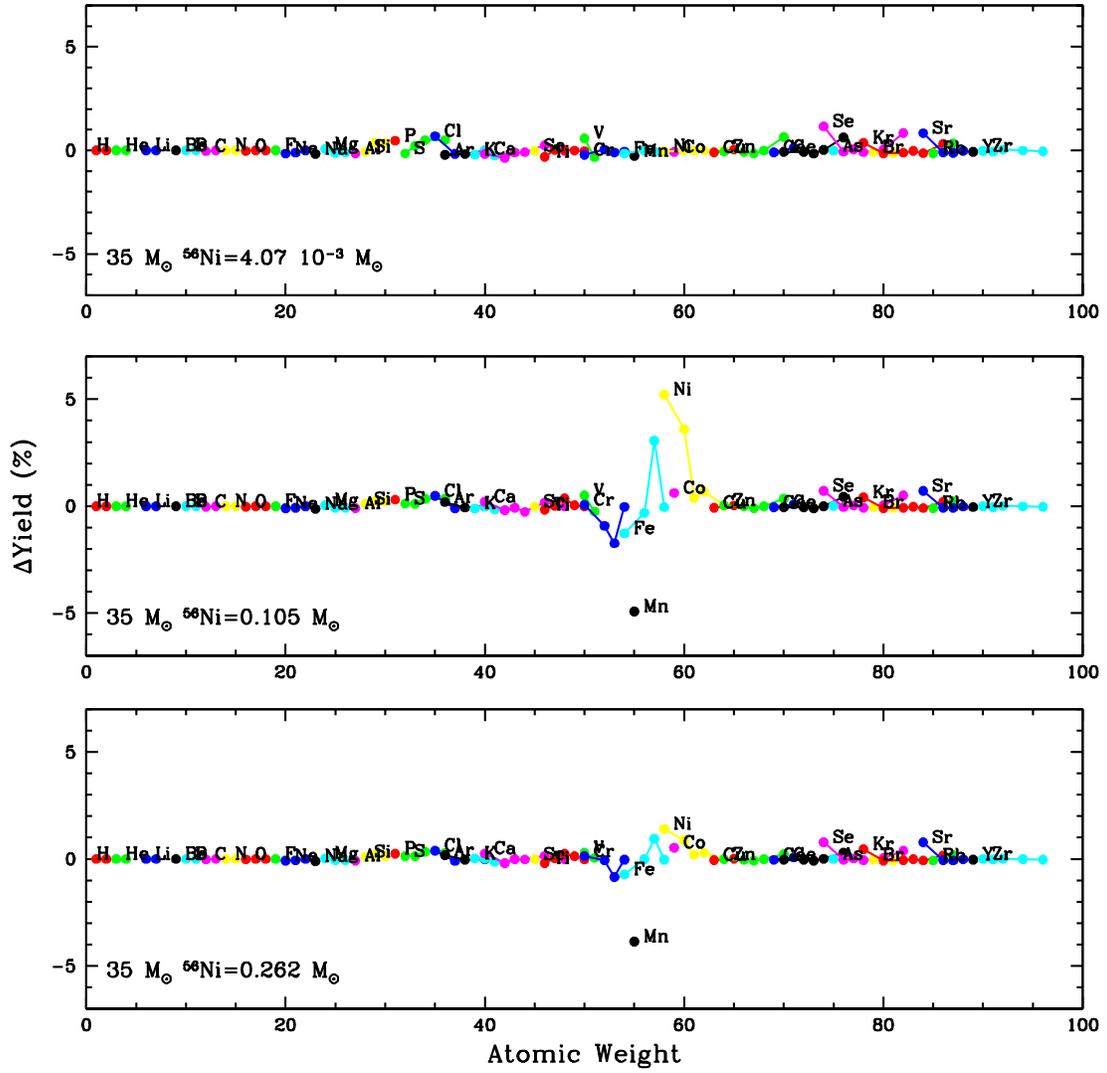}
\caption{Same as Figure \ref{yield15comp} but for the $\rm 35~M_\odot$ progenitor star.
\label{yield35comp}}
\end{figure}

To eject a prefixed amount of $\rm ^{56}Ni$ it is usually necessary to compute 
several hydro runs iterating on the initial velocity $v_{0}$. It would be much 
easier to run just an hydro model and obtain the given amount of $\rm ^{56}Ni$ 
(plus all the other yields) by imposing the mass cut a posteriori by hand. By 
means of the present runs we can check if this procedure is reliable or not. Let 
us check if, e.g., the yields provided by case 15B may be obtained from case 
15F. Hence let us impose in the 15F case a mass cut such that the yield of $\rm 
^{56}Ni$ is the same obtained in case 15B; let us call this hand made model 
15FB. The comparison between the "real" case 15B and the "fictitious" case 15FB is 
shown in Figure \ref{yield15comp} where the (percentage) differences between the 
two cases are shown in the upper panel. It is remarkable that the 
differences remain confined within 20\% for all the isotopes except that $\rm 
^{20,21}Ne$, $\rm ^{25}Mg$ and $\rm ^{46}Ca$. Also for these nuclei, however, 
the maximum difference never exceeds 50\%. To span a range of different mass 
cuts, the central panel in Figure \ref{yield15comp} shows the comparison between 
case 15C and a "fictitious" case 15FC while the lower panel shows the comparison 
between case 15D and 15FD. As expected, the discrepancies progressively reduce 
as the imposed mass cut gets closer to the real one. Figure \ref{yield35comp} 
shows a similar comparison for the $\rm 35~M_\odot$ model. In this case the 
maximum differences never exceeds 5\%.

\subsection{Comparison between RDA and Hydro calculations}

\begin{figure}
\figurenum{6}
\epsscale{1.0}
\plotone{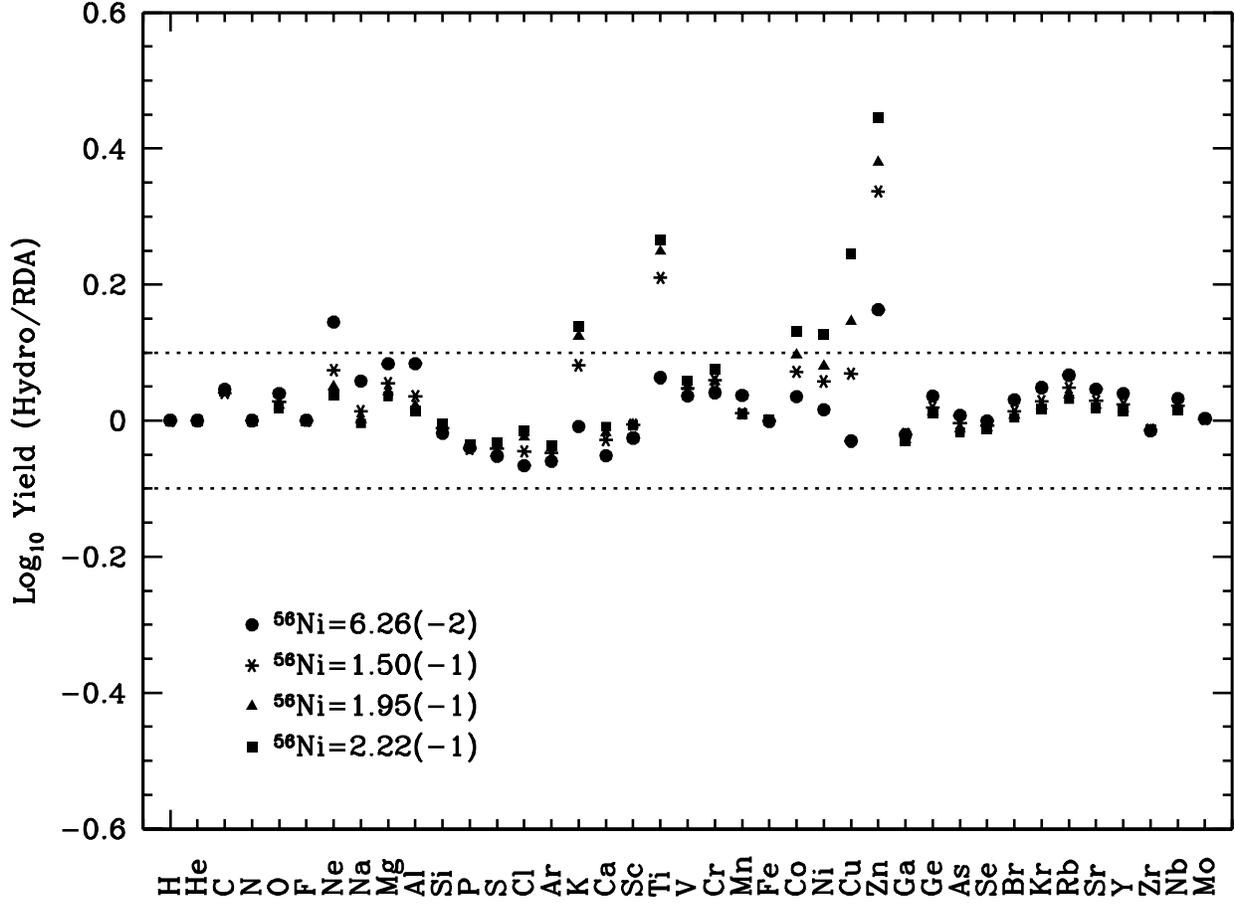}
\caption{Logarithmic ratio of the elemental yields obtained with the hydro code and with the RDA technique
for the $\rm 13~M_\odot$ for various choices of the $\rm ^{56}Ni$ ejected (reported in the legend).
\label{hydsurdaele}}
\end{figure}

\begin{figure}
\figurenum{7}
\epsscale{1.0}
\plotone{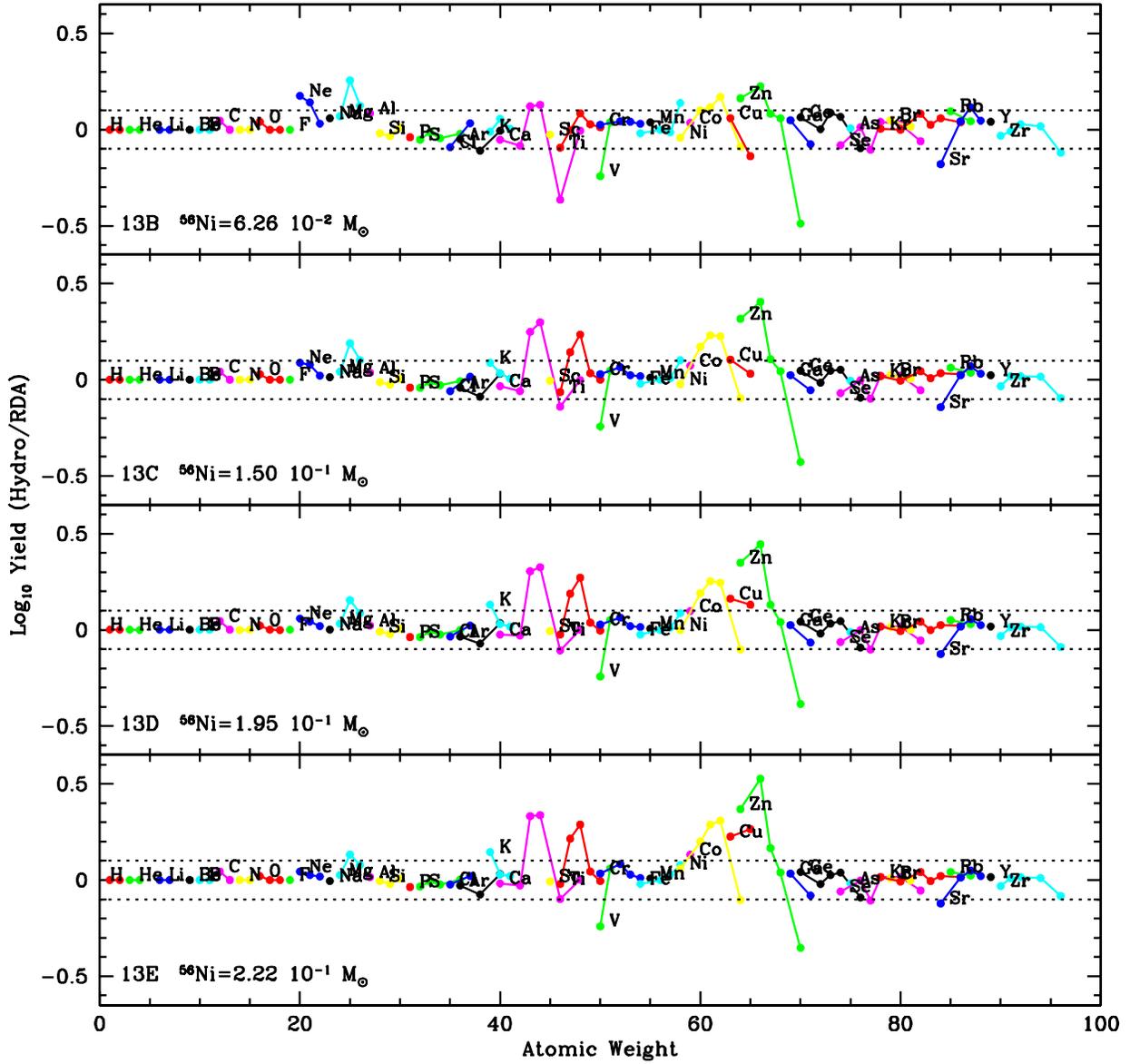}
\caption{Same as Figure \ref{hydsurdaele} but for the isotopic yields.
\label{hydsurdaiso}}
\end{figure}

It is worth comparing now the explosive yields obtained with the RDA technique 
and the ones obtained by means of the hydro code. Since, at variance with the 
RDA calculations, the hydro code does not take into account the time delay 
($\tau_{delay}$) between the core collapse and the explosion, we have 
recalculated the RDA explosions for all the models by imposing $\tau_{delay}=0$. 
Figure \ref{hydsurdaele} shows, for the $\rm 13~M_\odot$, the logarithmic ratio 
of the elemental yields obtained with the hydro code and the RDA technique for 
the same amount of ejected $\rm ^{56}Ni$. By the way, the elemental abundances 
are obtained by firstly decaying all the unstable nuclei to their parental 
stable neighbor and then summing up all the stable nuclei of each given element. 
Although the two techniques can be considered very different, it is surprising 
that the difference in the final yields are always confined within 0.1 dex for 
all the elements except for K, Ti, Co, Ni, Cu and Zn, i.e. the elements produced 
only (or mainly) in the zone that undergoes explosive Si burning with complete 
Si exhaustion. Note however that, even in the worst case, for these elements the 
differences between the two techniques remain confined within 0.5 dex, i.e. less 
than a factor of 3. A closer inspection to the temporal evolution of one 
meshpoint exposed to the complete explosive Si burning shows that the RDA 
technique overestimates both the peak temperature and the peak density and leads 
to a slower expansion compared to the hydro calculation. This is a well known 
feature of the radiation dominated shock approximation and it is also readily 
visible in, e.g., WW95 (their Figure 9). Although this occurrence can be 
generalized also to all the other zones, it mostly influences the region exposed 
to temperatures larger than 5 billion degrees, i.e. the ones characterized by 
the alpha rich freeze out. Such differences become less important as the mass of 
the progenitor star increases, hence the yields obtained with the two techniques 
become smaller as well. More specifically,the maximum difference is confined 
within 0.15 and 0.10 dex for the $\rm 25~M_\odot$ and the $\rm 35~M_\odot$ 
respectively. Figure \ref{hydsurdaiso} shows that even the isotopic differences 
between hydro and RDA calculations are nearly always confined within 0.1 dex 
with some notable exceptions, namely, $\rm ^{20}Ne$, $\rm ^{25}Mg$, $\rm 
^{39}K$, $\rm ^{43,44,46}Ca$, $\rm ^{47,48}Ti$, $\rm ^{50}V$, $\rm ^{59}Co$, 
$\rm ^{58,60,61,62}Ni$, $\rm ^{63,65}Cu$, $\rm ^{64,66,67,70}Zn$ and $\rm 
^{84}Sr$. $\rm ^{39}K$, $\rm ^{48}Ti$, $\rm ^{59}Co$, $\rm ^{58}Ni$, $\rm 
^{63,65}Cu$ and $\rm ^{64,66,67}Zn$ are the largest (if not the only) 
contributors to their respective elemental abundances and hence the possible 
origin of the differences obtained for these isotopes has already been discussed 
above. $\rm ^{47}Ti$, $\rm ^{43}Ca$ and $\rm ^{60,61,62}Ni$ are also produced 
only, or mainly, in the region of the alpha rich freeze out hence the differences 
can be readily understood. $\rm ^{20}Ne$ and $\rm ^{25}Mg$ are produced during 
the hydrostatic evolution in the carbon convective shell and then destroyed in 
all the zones shocked to a temperature larger than $\rm \sim 2\cdot 10^{9}~K$. 
As a consequence their final abundance will depend on the radial coordinate at 
which the peak temperature of $\rm \sim 2\cdot 10^{9}~K$ is reached. Such a 
critical radius is more external in the RDA compared to the hydro calculations. 
$\rm ^{46}Ca$, $\rm ^{50}V$, $\rm ^{70}Zn$ and $\rm ^{84}Sr$ are produced by the 
explosion; they show a gaussian-like profile with a maximum corresponding to a 
mass coordinate characterized by a given peak temperature that is $\rm \sim 
2\cdot 10^{9}~K$ for $\rm ^{46}Ca$, $\rm ^{50}V$ and $\rm ^{70}Zn$, and $\rm 
\sim 3\cdot 10^{9}~K$ for $\rm ^{84}Sr$. Both the mass location of this maximum 
and its width depend on the relation between the peak temperature and the 
interior mass. In general the RDA calculations tend to underproduce all of these 
three isotopes with respect to the hydro explosions, this effect being larger 
for lower explosion energies in the hydro models. $\rm ^{44}Ca$ is made by 
itself and by $\rm ^{44}Ti$ in proportions depending on the explosion energy -
$\rm ^{44}Ti$ is synthesized in the alpha rich freeze out while $\rm ^{44}Ca$ is 
produced by the explosion in more external zones (like, e.g., $\rm ^{46}Ca$). 
For highest energies most of the zone undergoing alpha rich freeze out is ejected 
and $\rm ^{44}Ca$ is completely made by $\rm ^{44}Ti$. For lower explosion 
energies a large amount of fall back occurs and $\rm ^{44}Ca$ is made only by 
itself. Both techniques give similar yields for $\rm ^{44}Ca$ for the lowest 
explosion energies while the hydro technique overproduces it $(\rm ^{44}Ti)$ 
because of a higher degree of alpha rich freeze out. All the differences discussed 
above progressively and significantly reduce as the mass of the progenitor 
increases.

\section{Discussion and Conclusions}

The yields provided by a generation of solar metallicity stars should be used, 
in principle, to interpret the ejecta of a core collapse supernova of similar 
initial metallicity as well as a part of a database of yields to be included 
into a galactic chemical evolutionary code. In principle there is no reason to 
require that the ejecta of a generation of solar metallicity stars preserve 
solar relative proportions because the Solar System distribution is the result 
of the cumulative contribution of many generations of stars of very different 
metallicities. Nonetheless, it is generally assumed the production factors (PFs) 
of a generation of solar metallicity stars to be essentially flat. This is the 
consequence of the (reasonable) assumption that the average metallicity Z grows 
slowly and continuously with respect to the evolutionary timescales of the stars 
that contribute to the environment enrichment; if this is true, the stars that 
mostly contribute to the abundances of the various nuclei at a given metallicity 
$\rm Z_0$ are those whose initial Z is quite close to $\rm Z_0$ (just to give an 
idea, the stars in the metallicity range $\rm Z_0>Z\ge Z_0/2$ contribute to 
roughly $50\%$ of the metallicity $\rm Z_0$, while stars in the range $\rm 
Z_0>Z\ge Z_0/10$ contribute to the metallicity $\rm Z_0$ for more than the 
$90\%$). Since the assumption mentioned above is certainly well verified at 
least for the massive stars, and very probably also for the intermediate mass 
ones, it follows that most of the Solar System distribution is the result (as a 
first approximation) of the ejecta of "quasi" solar metallicity stars. This is 
the reason why it is desirable that a generation of solar metallicity stars 
provides yields in roughly solar proportions or, in other words, that the PFs of 
the various nuclei remain roughly flat. Since Oxygen is produced only by the 
core collapse supernovae and it is also the most abundant element produced by 
these stars, it is convenient to use its PF as the one that better marks the 
overall increase of the average "metallicity" and to verify if the other nuclei 
follow or not its behavior. Arbitrarily we chose a factor of two as a suitable 
warning threshold in the sense that we will assume that all the nuclei whose PF 
falls within a factor of two of the Oxygen one are compatible with a flat 
distribution while those outside this compatibility range deserve a closer look-
up and may potentially constitute a problem.

\begin{figure}
\figurenum{8}
\epsscale{1.0}
\plotone{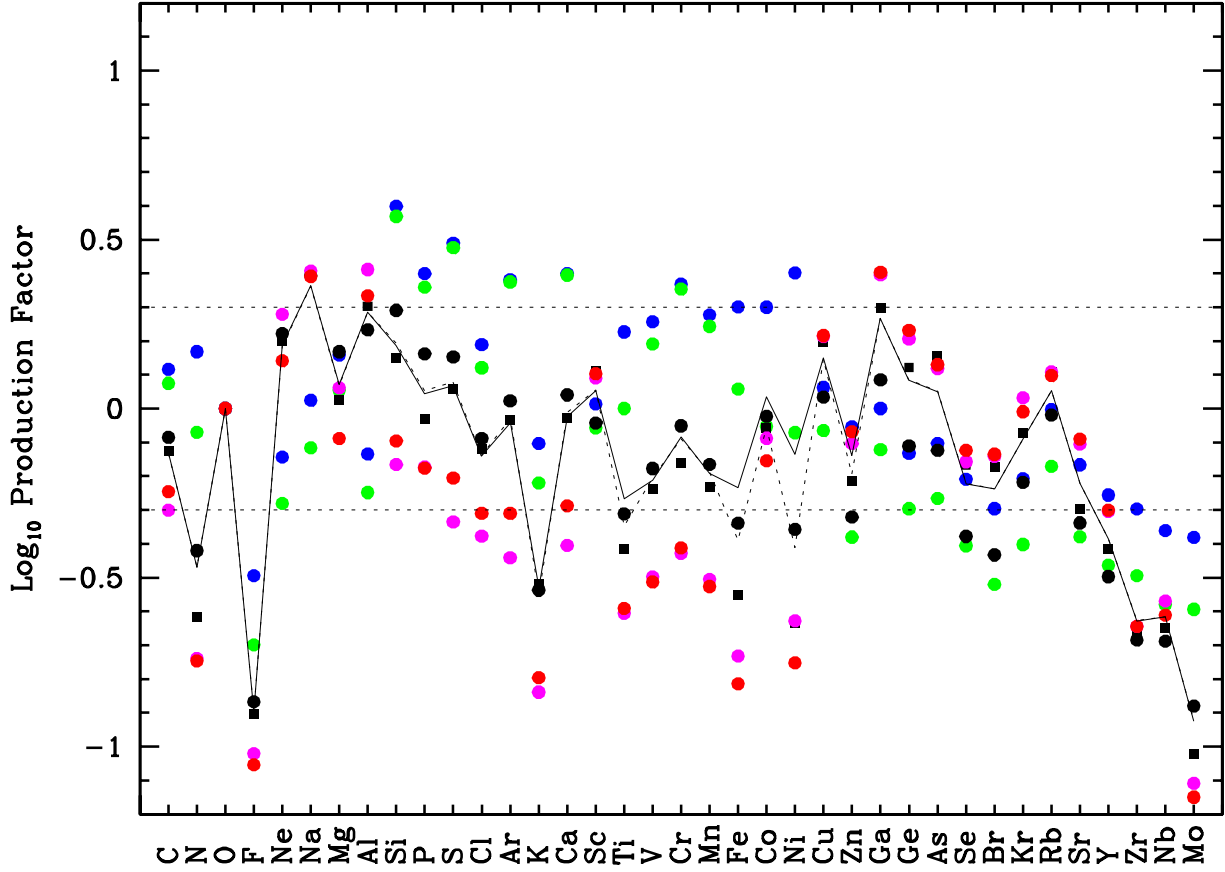}
\caption{ 
Production factors of all the elements from C to Mo. The symbols refer to the 6 
masses: $\rm 13~M_\odot$ ({\em blue dots}), $\rm 15~M_\odot$ ({\em green dots}),
$\rm 20~M_\odot$ ({\em black dots}), $\rm 25~M_\odot$ ({\em 
black squares}), $\rm 30~M_\odot$ ({\em magenta dots}), $\rm 35~M_\odot$ 
({\em red dots}). Note that all the PFs have been rescaled to allow the Oxygen PF 
to be at zero. The two lines refer to a generation of massive stars having a 
Salpeter mass function: the {\em dotted} line represents the case (Flat) in 
which all the core collapse supernovae are assumed to eject 0.05 $\rm M_\odot$ 
of $\rm ^{56}Ni$. The {\em solid} line is obtained assuming a given relation 
(Trend) between the mass cut and the progenitor mass (see text). 
\label{intele}} 
\end{figure}

Let us firstly concentrate on the elemental PFs (Figure \ref{intele}). The 
symbols refer to the 6 masses while the two lines refer to a generation of 
massive stars having a Salpeter mass function $(dn/dm\propto m^{- 2.35})$ but 
different choices for the mass cut. The dotted line (hereinafter case "Flat") 
refers to the case in which all the core collapse supernovae are assumed to 
eject 0.05 $\rm M_\odot$ of $\rm ^{56}Ni$ independently on the initial mass, 
while the solid one (hereinafter case "Trend") is obtained by assuming the 
following dependence of the mass cut (actually the amount of $\rm ^{56}Ni$ 
ejected) on the initial mass: $\rm 13~M_\odot$ $\rm (0.15~M_\odot~of~^{56}Ni)$, 
$\rm 15~M_\odot$ $\rm (0.10~ M_\odot~of~^{56}Ni)$, $\rm 20~M_\odot$ $\rm 
(0.08~M_\odot~of~^{56}Ni)$, $\rm 25~M_\odot$ $\rm (0.07~ M_\odot~of~^{56}Ni)$, 
$\rm 30~M_\odot$ $\rm (0.05~M_\odot~of~^{56}Ni)$, $\rm 35~M_\odot$ $\rm (0.05~ 
M_\odot~of~^{56}Ni)$. Since there is not a subset of models in our grid of 
explosions ejecting  the required $\rm ^{56}Ni)$ abundances as a function of the 
mass, we have obtained the desired models by choosing by hand the mass cut in 
models 13B, 15C, 20C, 25C, 30C and 35E. The hydro models shown in Figure 
\ref{intele} are those used to build up the "Flat" case and hence are those that 
eject 0.05 $\rm M_\odot$ of $\rm ^{56}Ni$. Since we are focusing on the relative 
scaling of the PFs with respect to that of the Oxygen, all the PFs have been 
renormalized by imposing $\rm Log(PF_{\rm Oxygen})=0$. This means that all the 
nuclei having a Log(PF) close to zero in Figure \ref{intele} preserve a scaled 
solar proportion with respect to the O. To facilitate the comparison with the 
yields provided by other authors we report in the first row of Table 11 our 
Oxygen PFs for the various masses while in rows 2 to 4 of the same Table we 
quote the yields given by \citet[RHHW02]{rau02}, \citet[TNH96]{tnh96} and 
\citet[WW95]{ww95}. The WW95 yields are all of the "A" variety while the TNH96 
ones are those computed with the original {\em Ye}. The first things worth noting in 
Figure \ref{intele} are the following: a) the yields produced by a generation of 
massive stars having a Salpeter mass function depend mainly on the yields of the 
masses between 20 and 25 $\rm M_\odot$ \citep{wzw78}; b) the only elements that 
vary significantly between the cases "Flat" (dotted) and "Trend" (solid) are Fe 
and Ni and, at a smaller extent, also Ti, Co and Zn; c) the majority of the 
elements from C to Sr have PFs compatible (in the sense mentioned above) with 
that of the Oxygen: the exceptions are N, F, Na, K, Ti, Fe and Ni.  N and F are 
largely underabundant (as expected) because our yields do not include neither 
the contributions from the intermediate mass stars (N) nor the neutrino induced 
reactions (F). The overproduction of Na (and partly of Al) is not totally 
unexpected because these two elements are "secondary", i.e. are elements whose 
production directly depends on the initial metallicity: the higher the initial 
metallicity the larger their final production. Hence their slight overproduction 
could simply indicate that the average metallicity to start with in order to get 
the solar system distribution should not be solar but slightly subsolar. By the 
way, similar results have been obtained by RHHW02. Let us also remark that the 
yields of these two elements strongly depend on the C abundance left by the 
central He burning \citep{ietal01} and in this paper we did not make any attempt 
to tune the $\rm ^{12}C(\alpha,\gamma)^{16}O$ reaction rate (nor the behavior of 
the convective core in the central He burning phase). K (produced mainly as $\rm 
^{39}K$ by both the explosive O burning and the complete explosive Si burning) 
is certainly significantly underproduced in our models but such a disagreement 
could be alleviated if the neutrino induced processes were efficient enough 
(WW95 found, for example, that the inclusion of the neutrino irradiation could 
increase the final abundance of $\rm ^{39}K$ up to a factor of two, see their 
Table 4). Ti, Fe and Ni depend significantly on the adopted mass cut and a 
changing from the "Flat" to the "Trend" case increases the yields of all these 
elements pushing them towards a closer scaled solar distribution. It must be 
noted, however, that the "Trend" case leaves less room for the Type Ia 
contribution because of the larger amount of Fe $\rm (^{56}Ni)$ produced (see 
below).  There are few  other things worth mentioning: since C has a PF close to 
zero, these models would predict that most of the C in the Solar System 
Distribution would come from the core collapse supernovae. The Iron peak 
elements that are synthesized only by the incomplete explosive Si burning (i.e. 
V, Cr and Mn) show PF's in good agreement with that of the O and do not show any 
dependence on the mass cut because the changes in the mass cut are confined (in 
the present test) to the region of the complete explosive Si burning. Sc does 
not show any dependence on the mass cut because in the range $\rm 20-25~M_\odot$ 
the yield of this element is dominated by the $\rm ^{45}Sc$ itself (that is 
produced in the central He burning and in the C burning shell) and not by the 
$\rm ^{45}Ca$ (that is synthesized by the complete explosive Si burning). Ti is 
produced mainly as $\rm ^{48}Cr$ and comes essentially from the complete and 
incomplete explosive Si burning. Co is produced directly as $\rm ^{59}Co$ (by 
the central He burning and the C burning shell) and as $\rm ^{59}Ni$ (by the 
complete explosive Si burning): in the "Flat" case the Co yield reflects mainly 
the hydrostatic production of $\rm ^{59}Co$ while in the "Trend" case there is 
also the contribution from the $\rm ^{59}Ni$. The Ni abundance is dominated by 
$\rm ^{58}Ni$ and $\rm ^{60}Ni$: the first of the two is destroyed by both the 
central He burning and by the convective C shell and is largely produced by both 
the complete and incomplete explosive Si burning; the second one is produced by 
the central He burning, the C and Ne explosive burnings and by the complete 
explosive Si burning. Cu is produced in similar amount as $\rm ^{63}Cu$, $\rm 
^{63}Ni$ and $\rm ^{65}Cu$; all these three isotopes are produced by the central 
He burning and the convective C shell. Zn is produced in comparable amounts as 
$\rm ^{64}Zn$, $\rm ^{66}Zn$ and $\rm ^{68}Zn$. The first of these three nuclei 
has a complex production history, since it is firstly largely produced by the 
central He burning but then it is significantly destroyed by the convective C-
shell; it is also produced in the layers experiencing an $\alpha$ rich freeze 
out in complete explosive Si burning. It is therefore clear that its final 
outcome will depend on a delicate balance between size of the convective core, 
outer border of the last convective C-shell and amount of fall back, if we limit 
ourself to the uncertainties connected with the physical evolution of the stars. 
In the present computations roughly 30\% of the $\rm ^{64}Zn$ yield comes from 
the complete explosive Si burning (for the mass cut adopted in the "Trend" 
case). The other two nuclei are, viceversa, produced by both the central He 
burning and the convective C shell. The PFs of the {\em s}-weak component, i.e. 
the elements from Ga to Sr, are compatible with those of the O and hence these 
models could provide the bulk of the abundances of these elements leaving a 
marginal role to the intermediate mass stars. The elements heavier than Sr are 
(correctly) largely underproduced because they come from the intermediate mass 
stars that are not included in the mass interval analyzed in this paper.

\begin{figure}
\figurenum{9}
\epsscale{1.0}
\plotone{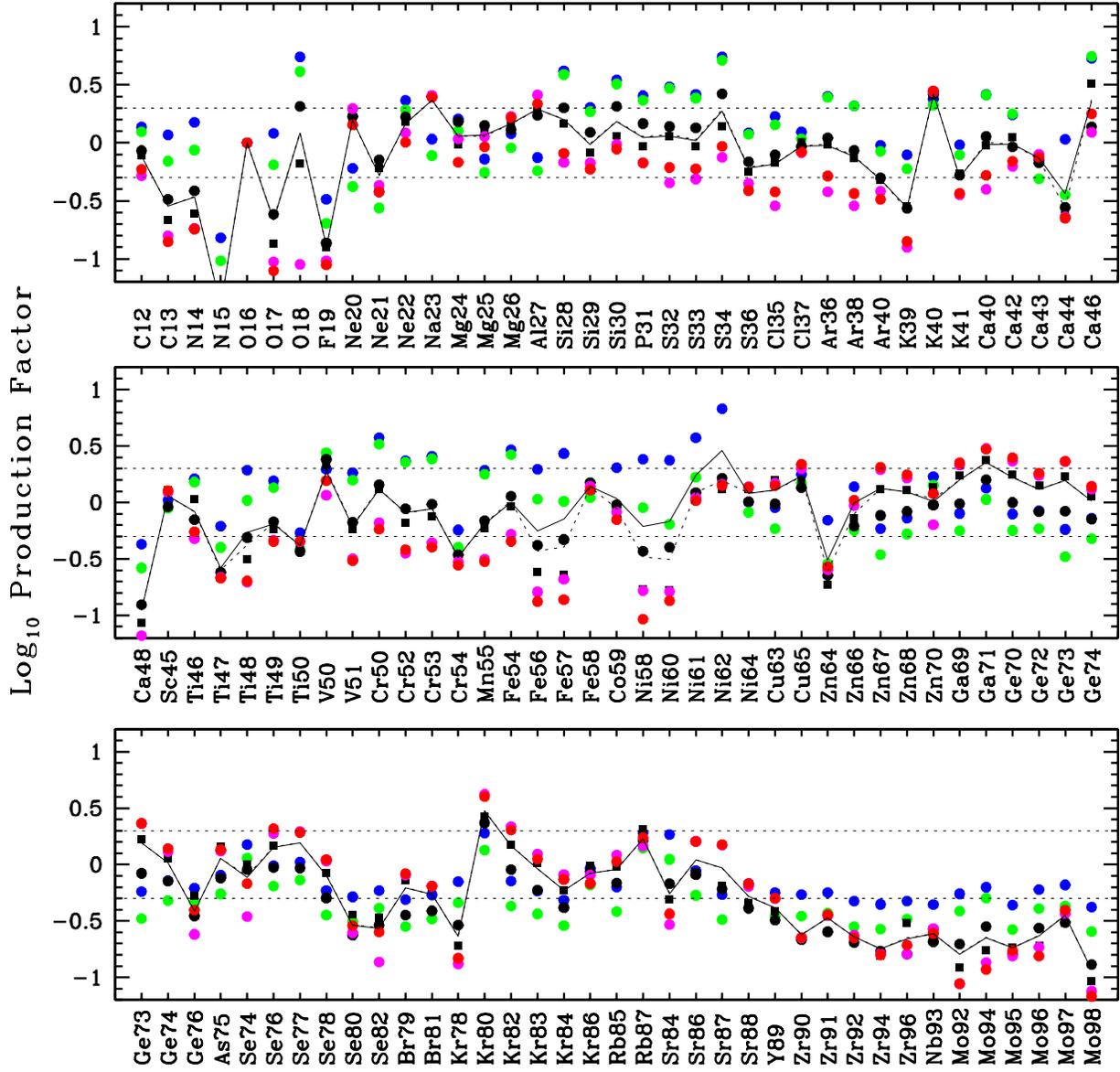}
\caption{Same as Figure 8 but for the isotopes.
\label{intiso}}
\end{figure}

Let us now turn to the isotopic PFs. Figure \ref{intiso} is the isotopic version 
of Figure \ref{intele}. As one could expect the main differences between the two 
integrated PFs concern the nuclei synthesized by the complete explosive Si 
burning, i.e. $\rm ^{44}Ti$ $\rm (^{44}Ca)$, $\rm ^{48}Cr$ $\rm (^{48}Ti)$, $\rm 
^{56}Ni$ $\rm (^{56}Fe)$, $\rm ^{57}Ni$ $\rm (^{57}Fe)$, $\rm ^{59}Ni$ $\rm 
(^{59}Co)$, $\rm ^{58}Ni$, $\rm ^{60}Ni$, $\rm ^{61}Ni$, $\rm ^{62}Ni$, $\rm 
^{64}Zn$ and $\rm ^{66}Zn$. Most of the nuclei in the range $\rm ^{12}C$ - $\rm 
^{64}Ni$ are compatible with a flat distribution of the PFs, with 17 nuclei 
being significantly underproduced and 4 nuclei largely overproduced. While the 
nuclei underproduced by the massive stars could come from other astrophysical 
sites (like, e.g., the $\rm ^{14}N$ that is very probably produced by the 
intermediate mass stars), it is clear that it is much more difficult to account 
for the nuclei overproduced; our models predict only 4 nuclei overabundant by 
more than a factor of two with respect to the O: they are $\rm ^{23}Na$, $\rm 
^{40}K$, $\rm ^{46}Ca$ and $\rm ^{62}Ni$. The PF of the $\rm ^{23}Na$ has been 
already discussed above. $\rm ^{40}K$ is an unstable nucleus with an half life 
of 1.25 Gyr; since the age of the Sun is 4.6 Gyr, the present solar abundance of 
$\rm ^{40}K$ must be increased by almost a factor of 10 to obtain its abundance 
at the moment of the solar system formation. If we apply such a correction to 
the solar abundance of $\rm ^{40}K$, its PF lowers to $\rm PF(^{40}K)\sim-0.4$ 
becoming now significantly underproduced with respect to O. $\rm ^{46}Ca$ is 
produced by both the convective C-shell and the Ne explosive burning: in both 
cases the fuel that feeds its production is the neutron flux that, in turn, 
largely depends on the amount of $\rm ^{22}Ne$ available and hence on the 
initial metallicity Z; this could mean that, as already suggested for Na, the 
schematic assumption that the solar system distribution may be attributed to the 
ejecta of solar metallicity stars fails significantly in this case. $\rm 
^{62}Ni$ is produced by the complete explosive Si burning and hence its PF 
largely depends on the adopted mass cut: it is indeed overproduced only in the 
"Trend" case.

Also the PFs of the nuclei between Cu and Sr are in reasonable agreement with that 
of the O, confirming the largely accepted idea that these elements are 
significantly produced by massive stars. 7 isotopes in this range scatter by more 
than a factor of 2 relative to O: $\rm ^{64}Zn$, $\rm ^{76}Ge$, $\rm ^{80}Se$, $\rm 
^{82}Se$, $\rm ^{78}Kr$, $\rm ^{71}Ga$ and $\rm ^{80}Kr$. The first 5 are 
underproduced while the last two are overproduced. $\rm ^{76}Ge$ and $\rm ^{82}Se$ 
have a strong {\em r}-process contribution, $\rm ^{78}Kr$ has probably a strong 
{\em p}-process contribution while the abundance of $\rm ^{80}Kr$ strongly depends 
on the lifetime of $\rm ^{79}Se$ that is still fairly uncertain for temperatures 
larger than $\rm  7\cdot 10^{8}~K$. It has been long recognized the block of 
elements between Cu and the neutron closure shell at N=50 cannot be produced mainly 
by the main {\em s}- component but that they require the existence of a weak {\em 
s}-component (early attributed to the central He burning of massive stars). However 
most of the matter exposed to the central He burning phase will be also exposed to 
further burnings, i.e. the C-burning and the passage of the shock wave. The 
contribution of these three phases to the yields are reported in Table 11. A clear 
trend exists: the main production site of most of these nuclei progressively shifts 
from the explosive burnings towards the C convective shell and central He burning 
as the mass of the progenitor increases. In particular, in the 13 and 15 $\rm 
M_\odot$ models all these isotopes are produced by the explosion itself. Viceversa, 
in the 20 - 35 $\rm M_\odot$ range most of these nuclei are produced by either the 
C-burning shell and/or the central He burning, the relative contribution 
privileging  the C- burning shell for the smaller masses and the central He burning 
for the larger ones. The negative percentages that appear in Table 11 refer to the 
percentage {\it destroyed} by that evolutionary phase: for example, in the $\rm 
30~M_\odot$, the $\rm C_{sh}$ burning destroys half of the $\rm ^{64}Zn$ produced 
by the central He burning.

\begin{figure}
\figurenum{10}
\epsscale{1.0}
\plotone{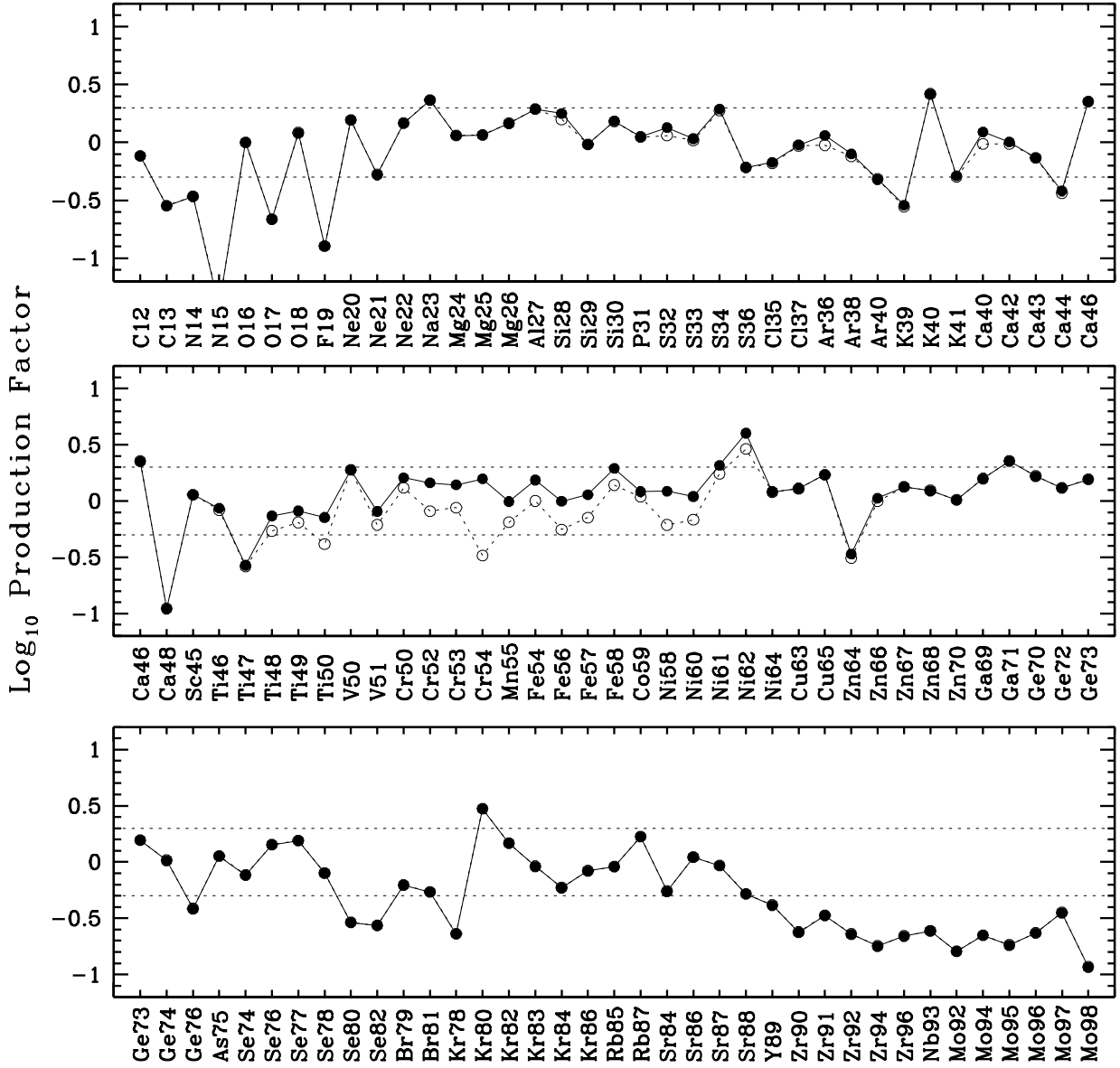}
\caption{Production factors of all the isotopes from $\rm ^{12}C$ to $\rm ^{98}Mb$. 
The two lines refer to the "Trend" case: the {\em open} dots (connected by the dashed line) show the 
Production Factors without the Type Ia contribution
while the {\em filled} dots (connected by the solid line) includes a 12\% Type Ia contribution.
\label{int_totali}}
\end{figure}

As for the contribution of the Type Ia supernovae to the Solar System 
distribution, we have tentatively adopted the yields provided by 
\cite{Iwetal99}. In this paper a variety of models spanning different values of 
central ignition density, flame speed and deflagration to detonation critical 
density have been presented. By evaluating the pros and the cons of their 
various models, \cite{Iwetal99} concluded that the CS15DD2 is the one that 
better reproduces the overall properties of type Ia supernovae (we refer the 
reader to the quoted paper for a deeper insight in the properties of these 
models). Though we have adopted the CS15DD2 model as the typical one describing 
the yields provided by the type Ia supernovae, even the adoption of the other 
models presented in that paper would not change the main results discussed 
below. The percentage frequency of Type Ia supernovae relative to the core 
collapse ones has been fixed by imposing $\rm PF_{Fe}$ = $\rm PF_{O}$. In the 
"Trend" case we find that this contribution is 12\% while in the "Flat" case it 
increases to 16\%. Figure \ref{int_totali} shows the quoted PF's with (filled 
dots connected by a solid line) and without (open dots connected by a dashed 
line) the Type Ia contribution for the "Trend" case. From this figure it is 
quite clear that the Type Ia supernovae contribute only to the Solar System 
abundances of the nuclei in the range Ti-Ni. Moreover, the inclusion of the Type 
Ia supernovae brings both $\rm ^{50}Ti$ and $\rm ^{54}Cr$ within the 
compatibility band that now includes all the nuclei in the range Ti-Ni, with the 
exception of $\rm ^{62}Ni$ whose discrepancy is, viceversa, slightly increased. 
Note that, at variance with current believing, Type Ia supernova do not 
contribute appreciably to the synthesis of $\rm ^{48}Ca$ which hence remains 
largely unexplained. Our models already predict $\rm 2.48 \times 10^{-6}$ $\rm 
M_\odot$ of $\rm ^{48}Ca$ per Type II supernovae while the Type Ia supernovae 
\citep{Iwetal99} produce at most $\rm 1.64 \times 10^{-6}$ $\rm M_\odot$ of $\rm 
^{48}Ca$ per Type Ia explosion: since there is room just for only 12\% of Type 
Ia supernova, it is readily understandable why these Type Ia yields do not 
change appreciably the total yield of this nucleus.

As a last comment, a changing of the slope of the Salpeter mass function from 
$\rm \alpha$=2.35 to $\rm \alpha$=0 would change the percentage of the Type Ia 
supernovae relative to the core collapse ones from 12\% to 28\%; such a changing 
would not alter significantly the present results.

\acknowledgements

We warmly thank Tony Mezzacappa for very helpful discussions, Franz K\"appeler 
for useful discussions about the lifetime of $\rm ^{79}Se$ and Roberto Gallino 
for having pushed us to start these computations. We also thank Brad 
Gibson and John Lattanzio for their kind hospitality in Melbourne.

\clearpage

\clearpage

\clearpage

\clearpage

\begin{deluxetable}{lrrlrr}\label{tabnet}
\tabletypesize{\scriptsize}
\tablewidth{0pt}
\tablecaption{Nuclear network adopted in the present calculations}
\tablehead{
\colhead{Element} & \colhead{$\rm A_{\rm min}$} & \colhead{$\rm A_{\rm max}$} &
\colhead{Element} & \colhead{$\rm A_{\rm min}$} & \colhead{$\rm A_{\rm max}$}
}
\startdata
n........  &   1   &   1   & Ti.......  &  44   &  51   \\
H........  &   1   &   3   & V........  &  45   &  52   \\
He.......  &   3   &   4   & Cr.......  &  48   &  55   \\
Li.......  &   6   &   7   & Mn.......  &  51   &  57   \\
Be.......  &   7   &  10   & Fe.......  &  52   &  61   \\
B........  &  10   &  11   & Co.......  &  55   &  61   \\
C........  &  12   &  14   & Ni.......  &  56   &  65   \\
N........  &  13   &  16   & Cu.......  &  57   &  66   \\
O........  &  15   &  19   & Zn.......  &  64   &  71   \\
F........  &  17   &  20   & Ga.......  &  65   &  72   \\
Ne.......  &  20   &  23   & Ge.......  &  70   &  77   \\
Na.......  &  21   &  24   & As.......  &  71   &  77   \\
Mg.......  &  23   &  27   & Se.......  &  74   &  83   \\
Al.......  &  25   &  28   & Br.......  &  75   &  83   \\
Si.......  &  27   &  32   & Kr.......  &  78   &  87   \\
P........  &  29   &  34   & Rb.......  &  79   &  88   \\
S........  &  31   &  37   & Sr.......  &  84   &  91   \\
Cl.......  &  33   &  38   & Y........  &  85   &  91   \\
Ar.......  &  36   &  41   & Zr.......  &  90   &  97   \\
K........  &  37   &  42   & Nb.......  &  91   &  97   \\
Ca.......  &  40   &  49   & Mo.......  &  92   &  98   \\
Sc.......  &  41   &  49   &            &       &       \\
\enddata
\end{deluxetable}

\begin{deluxetable}{lcccccc}
\tabletypesize{\scriptsize}
\tablewidth{15cm}
\tablecaption{Selected quantities for $\rm 13~M_\odot$, $\rm 15~M_\odot$, $\rm 20~M_\odot$, $\rm 25~M_\odot$, $\rm 30~M_\odot$ and $\rm 35~M_\odot$ PSN models.}
\tablehead{
\colhead{}   &
\colhead{$\rm 13~M_\odot$}    &
\colhead{$\rm 15~M_\odot$}    &
\colhead{$\rm 20~M_\odot$}    &
\colhead{$\rm 25~M_\odot$}    &
\colhead{$\rm 30~M_\odot$}    &
\colhead{$\rm 35~M_\odot$}    
} 
\startdata  
\cutinhead{H Burning}
 Central H burning lifetime (yr)                    &   1.26(7)   &   1.03(7)    &   7.26(6)    &    5.79(6)  &   4.90(6)    &    4.34(6)  \\
 Max size of the conv. core ($\rm M_\odot$)         &    4.83     &    6.01      &   9.22       &   12.80     &  16.40       &    20.08    \\   
 $\rm M_{\rm He~core}^{\rm final}$ ($\rm M_\odot$)    &    3.61     &    4.29      &   6.49       &    8.82     &  11.23       &   13.16     \\  
 $\rm R_{\rm He~core}^{\rm final}$ ($\rm R_\odot$)    &    2.14     &    8.71(-1)  &   6.58(-1)   &    6.18(-1) &   5.90(-1)   &    5.27(-1) \\  
\cutinhead{Convective Envelope}                                                      
 Final location of the conv. env. ($\rm M_\odot$)   &    3.65     &    4.43      &   6.77       &    9.21     &  11.89       &   13.47     \\  
 $\rm Y_{surface}$                                             &   3.07(-1)  &   3.09(-1)  &   3.16(-1)   &   3.24(-1)  &   3.30(-1)   &   3.52(-1)   \\
 $\rm ^{12}C_{surface}$                                        &   1.94(-3)  &   2.01(-3)  &   2.04(-3)   &   2.05(-3)  &   2.05(-3)   &   1.97(-3)   \\
 $\rm ^{13}C_{surface}$                                        &   1.21(-4)  &   1.21(-4)  &   1.25(-4)   &   1.28(-4)  &   1.29(-4)   &   1.25(-4)   \\
\enddata        
\end{deluxetable}

\begin{deluxetable}{lcllllllllll}
\tabletypesize{\scriptsize}
\tablewidth{16cm}
\tablecaption{Selected properties of all the exploded models}
\tablehead{
\colhead{Mass} &
\colhead{Label} &
\colhead{$v_{0}$}    &
\colhead{$t_{\rm CO}$}    &
\colhead{$t_{\rm sfb}$}    &
\colhead{$t_{\rm efb}$}    &
\colhead{$t_{\rm srv}$}    &
\colhead{$t_{\rm break}$}    &
\colhead{$t_{\rm erv}$}    &
\colhead{$E_{\rm kin}$}    &
\colhead{$\rm M_{\rm rem}$}    &
\colhead{$\rm ^{56}Ni$}  \\
\colhead{($\rm M_\odot$)}    &
   &
\colhead{cm/s}    &
\colhead{s}    &
\colhead{s}    &
\colhead{s}    &
\colhead{s}    &
\colhead{s}    &
\colhead{s}    &
\colhead{foe}    &
\colhead{($\rm M_\odot$)}    &
\colhead{($\rm M_\odot$)}  
} 
\startdata  
13 & A & 1.25000(9)&  0.963  &   7.30(-1)  &   1.90(5)   &      1.56(3)   &    1.18(5)   &    6.12(5)  &   0.450  &    1.56 &  2.72(-2)    \\
13 & B & 1.29000(9)&  0.875  &   8.75(-1)  &   1.84(5)   &      1.48(3)   &    1.13(5)   &    5.38(5)  &   0.500  &    1.52 &  6.26(-2)    \\
13 & C & 1.42000(9)&  0.794  &   1.66(+0)  &   1.03(3)   &      1.27(3)   &    9.64(4)   &    1.78(5)  &   0.800  &    1.41 &  1.50(-1)    \\
13 & D & 1.47000(9)&  0.761  &   2.57(+0)  &   8.41(2)   &      1.20(3)   &    9.10(4)   &    1.47(5)  &   0.960  &    1.36 &  1.95(-1)    \\
13 & E & 1.50000(9)&  0.761  &   3.26(+0)  &   1.82(3)   &      1.20(3)   &    8.82(4)   &    1.08(5)  &   1.090  &    1.24 &  2.22(-1)    \\
13 & F & 1.53500(9)&  0.757  &   5.31(+0)  &   4.57(2)   &      1.12(3)   &    8.49(4)   &    8.44(4)  &   1.078  &    1.32 &  2.53(-1)    \\
13 & G & 1.55000(9)&  0.720  &   6.75(+0)  &   2.30(2)   &      1.11(3)   &    8.36(4)   &    1.01(5)  &   1.158  &    1.21 &  2.54(-1)    \\
15 & A & 1.22000(9)&  1.430  &   6.36(-1)  &   5.93(6)   &      3.09(3)   &    1.50(5)   &    7.00(5)  &   0.387  &    1.99 &  7.57(-15)   \\
\enddata         
\end{deluxetable}

\begin{deluxetable}{lrrrrrrrrrr}
\tabletypesize{\scriptsize}
\tablewidth{0cm}
\tablecaption{Explosive yields relative to the various explosions of the $\rm 13~M_\odot$ model.
The unstable isotopes have been decayed into their stable form}
\tablehead{
\colhead{Isotope} &
\colhead{13A} &
\colhead{13B} &
\colhead{13C} &
\colhead{13D} &
\colhead{13E} &
\colhead{13F} &
\colhead{13G}
} 
\startdata  
$\rm E_{\rm kin}$ &    0.450  &     0.500  &      0.800 &      0.960  &     1.078 &      1.090  &      1.158  \\
$\rm M_{\rm  ej}$ &    11.44  &     11.48  &      11.59 &      11.64  &     11.76 &      11.68  &      11.79  \\  
$\rm M_{\rm rem}$ &     1.56  &      1.52  &       1.41 &       1.36  &      1.24 &       1.32  &       1.21  \\
$\rm ^{56}Ni$     & 2.720(-2) &  6.260(-2) &   1.505(-1)&   1.946(-1) &  2.529(-1)&   2.215(-1) &   2.536(-1) \\
$\rm ^{1}H  $     & 6.23      &  6.23      &   6.23     &   6.23      &  6.23     &   6.23      &   6.23      \\
$\rm ^{2}H  $     & 1.54(-16) &  1.54(-16) &   1.54(-16)&   1.54(-16) &  1.54(-16)&   1.54(-16) &   1.54(-16) \\
$\rm ^{3}He $     & 4.86(-04) &  4.86(-04) &   4.86(-04)&   4.86(-04) &  4.86(-04)&   4.86(-04) &   4.86(-04) \\
$\rm ^{4}He $     & 4.37      &  4.37      &   4.37     &   4.37      &  4.37     &   4.37      &   4.37      \\
\enddata         
\end{deluxetable}

\begin{deluxetable}{lrrrrrrrrr}
\tabletypesize{\scriptsize}
\tablewidth{0cm}
\tablecaption{Explosive yields relative to the various explosions of the $\rm 15~M_\odot$ model.
The unstable isotopes have been decayed into their stable form}
\tablehead{
\colhead{Isotope} &
\colhead{15A} &
\colhead{15B} &
\colhead{15C} &
\colhead{15D} &
\colhead{15E} &
\colhead{15F}
} 
\startdata  
$\rm E_{\rm kin}$ &    0.387  &      0.600 &     0.862 &      1.050  &      1.090 &      1.275   \\
$\rm M_{\rm  ej}$ &    13.01  &      13.28 &     13.37 &      13.49  &      13.60 &      13.83   \\  
$\rm M_{\rm rem}$ &     1.99  &       1.72 &      1.63 &       1.51  &       1.40 &       1.17   \\
$\rm ^{56}Ni$     &7.570(-15) &  8.280(-3) & 9.030(-2) &  1.911(-1)  &  2.782(-1) &  2.873(-1)   \\
$\rm ^{1}H  $     &     7.03  &      7.03  &     7.03  &      7.03   &      7.03  &      7.03    \\
$\rm ^{2}H  $     & 1.64(-16) &  1.64(-16) & 1.64(-16) &  1.64(-16)  &  1.64(-16) &  1.64(-16)   \\
$\rm ^{3}He $     & 5.65(-04) &  5.65(-04) & 5.65(-04) &  5.65(-04)  &  5.65(-04) &  5.65(-04)   \\
$\rm ^{4}He $     &     4.93  &      4.93  &     4.93  &      4.93   &      4.93  &      4.93    \\
\enddata         
\end{deluxetable}

\begin{deluxetable}{lrrrrrrrrrr}
\tabletypesize{\scriptsize}
\tablewidth{0cm}
\tablecaption{Explosive yields relative to the various explosions of the $\rm 20~M_\odot$ model.
The unstable isotopes have been decayed into their stable form}
\tablehead{
\colhead{Isotope} &
\colhead{20B} &
\colhead{20C} &
\colhead{20D} &
\colhead{20E} &
\colhead{20F} &
\colhead{20G} &
\colhead{20H}
} 
\startdata  
$\rm E_{\rm kin}$ &    0.873  &     0.950  &     0.997  &     1.022   &     1.032 &     1.041   &     1.048   \\
$\rm M_{\rm  ej}$ &    18.03  &     18.25  &     18.38  &     18.47   &     18.51 &     18.54   &     18.62   \\  
$\rm M_{\rm rem}$ &     1.97  &      1.75  &      1.62  &      1.53   &      1.49 &      1.46   &      1.38   \\
$\rm ^{56}Ni$     &4.330(-7)  & 7.720(-2)  & 1.758(-1)  & 2.412(-1)   & 2.720(-1) & 2.952(-1)   & 3.254(-1)   \\
$\rm ^{1}H  $     &      8.72 &      8.72  &      8.72  &      8.72   &      8.72 &      8.72   &      8.72   \\
$\rm ^{2}H  $     & 1.90(-16) & 1.90(-16)  & 1.90(-16)  & 1.90(-16)   & 1.90(-16) & 1.90(-16)   & 1.90(-16)   \\
$\rm ^{3}He $     & 6.92(-04) & 6.92(-04)  & 6.92(-04)  & 6.92(-04)   & 6.92(-04) & 6.92(-04)   & 6.92(-04)   \\
$\rm ^{4}He $     &      6.45 &      6.45  &      6.45  &      6.45   &      6.45 &      6.45   &      6.45   \\
\enddata         
\end{deluxetable}

\begin{deluxetable}{lrrrrrrrrr}
\tabletypesize{\scriptsize}
\tablewidth{0cm}
\tablecaption{Explosive yields relative to the various explosions of the $\rm 25~M_\odot$ model.
The unstable isotopes have been decayed into their stable form}
\tablehead{
\colhead{Isotope} &
\colhead{25A} &
\colhead{25B} &
\colhead{25C} &
\colhead{25D} &
\colhead{25E} &
\colhead{25F}
} 
\startdata  
$\rm E_{\rm kin}$ &     1.110 &     1.144  &     1.170  &     1.220   &     1.250 &     1.263    \\
$\rm M_{\rm  ej}$ &     22.99 &     23.11  &     23.20  &     23.30   &     23.44 &     23.77    \\  
$\rm M_{\rm rem}$ &      2.01 &      1.89  &      1.80  &      1.70   &      1.56 &      1.23    \\
$\rm ^{56}Ni$     & 2.360(-3) & 3.390(-2)  & 1.092(-1)  & 1.887(-1)   & 2.950(-1) & 3.034(-1)    \\
$\rm ^{1}H  $     & 1.03(+01) & 1.03(+01)  & 1.03(+01)  & 1.03(+01)   & 1.03(+01) & 1.03(+01)    \\
$\rm ^{2}H  $     & 2.16(-16) & 2.16(-16)  & 2.16(-16)  & 2.16(-16)   & 2.16(-16) & 2.16(-16)    \\
$\rm ^{3}He $     & 8.14(-04) & 8.14(-04)  & 8.14(-04)  & 8.14(-04)   & 8.14(-04) & 8.14(-04)    \\
$\rm ^{4}He $     &      7.84 &      7.84  &      7.84  &      7.84   &      7.84 &      7.84    \\
\enddata         
\end{deluxetable}

\begin{deluxetable}{lrrrrrrrrrr}
\tabletypesize{\scriptsize}
\tablewidth{0cm}
\tablecaption{Explosive yields relative to the various explosions of the $\rm 30~M_\odot$ model.
The unstable isotopes have been decayed into their stable form}
\tablehead{
\colhead{Isotope} &
\colhead{30A} &
\colhead{30B} &
\colhead{30C} &
\colhead{30D} &
\colhead{30E} &
\colhead{30F} &
\colhead{30G}
} 
\startdata  
$\rm E_{\rm kin}$ &     0.859 &     1.037  &     1.031  &     1.032   &     1.036 &     1.038   &     1.039   \\
$\rm M_{\rm  ej}$ &     27.76 &     28.15  &     28.28  &     28.38   &     28.46 &     28.50   &     28.53   \\  
$\rm M_{\rm rem}$ &      2.24 &      1.85  &      1.72  &      1.62   &      1.54 &      1.50   &      1.47   \\
$\rm ^{56}Ni$     &5.000(-15) & 4.700(-3)  & 7.710(-3)  & 1.503(-1)   & 2.082(-1) & 2.427(-1)   & 2.535(-1)   \\
$\rm ^{1}H  $     & 1.18(+01) & 1.18(+01)  & 1.18(+01)  & 1.18(+01)   & 1.18(+01) & 1.18(+01)   & 1.18(+01)   \\
$\rm ^{2}H  $     & 2.41(-16) & 2.41(-16)  & 2.41(-16)  & 2.41(-16)   & 2.41(-16) & 2.41(-16)   & 2.41(-16)   \\
$\rm ^{3}He $     & 9.31(-04) & 9.31(-04)  & 9.31(-04)  & 9.31(-04)   & 9.31(-04) & 9.31(-04)   & 9.31(-04)   \\
$\rm ^{4}He $     &      9.17 &      9.17  &      9.17  &      9.17   &      9.17 &      9.17   &      9.17   \\
\enddata         
\end{deluxetable}

\begin{deluxetable}{lrrrrrrrrrr}
\tabletypesize{\scriptsize}
\tablewidth{0cm}
\tablecaption{Explosive yields relative to the various explosions of the $\rm 35~M_\odot$ model.
The unstable isotopes have been decayed into their stable form}
\tablehead{
\colhead{Isotope} &
\colhead{35B} &
\colhead{35D} &
\colhead{35E} &
\colhead{35G} &
\colhead{35H} &
\colhead{35I} &
\colhead{35J}
} 
\startdata  
$\rm E_{\rm kin}$ &     1.360 &     1.520  &     1.610  &     1.633   &     1.668 &     1.680   &     1.740   \\
$\rm M_{\rm  ej}$ &     32.18 &     32.72  &     32.92  &     32.96   &     33.06 &     33.12   &     33.94   \\  
$\rm M_{\rm rem}$ &      2.82 &      2.28  &      2.08  &      2.04   &      1.94 &      1.88   &      1.06   \\
$\rm ^{56}Ni$     & 3.57(-14) & 4.070(-3)  & 1.054(-1)  & 1.399(-1)   & 2.169(-1) & 2.620(-1)   & 4.831(-1)   \\
$\rm ^{1}H  $     & 1.34(+01) & 1.34(+01)  & 1.34(+01)  & 1.34(+01)   & 1.34(+01) & 1.34(+01)   & 1.34(+01)   \\
$\rm ^{2}H  $     & 2.50(-16) & 2.50(-16)  & 2.50(-16)  & 2.50(-16)   & 2.50(-16) & 2.50(-16)   & 2.50(-16)   \\
$\rm ^{3}He $     & 1.04(-03) & 1.04(-03)  & 1.04(-03)  & 1.04(-03)   & 1.04(-03) & 1.04(-03)   & 1.04(-03)   \\
$\rm ^{4}He $     & 1.07(+01) & 1.07(+01)  & 1.07(+01)  & 1.07(+01)   & 1.07(+01) & 1.07(+01)   & 1.07(+01)   \\
\enddata                                                                                                            
\end{deluxetable}

\begin{deluxetable}{lrrrrrrrrrr}
\tabletypesize{\scriptsize}
\tablewidth{0cm}
\tablecaption{Ejected masses at $\rm t=2.5\cdot10^{4}~s$ of all the radioactive isotopes}
\tablehead{
\colhead{Isotope} &
\colhead{(1)} &
\colhead{(2)} &
\colhead{(3)} &
\colhead{(4)} &
\colhead{(5)} &
\colhead{(6)} &
\colhead{(7)}
} 
\startdata  
\cutinhead{$\rm 13~M_\odot$}
$\rm model  $ &     13A     &     13B     &     13C    &      13D    &      13E    &      13F    &      13G     \\ 
$\rm E_{kin}$ &      0.450  &      0.500  &      0.800 &       0.960 &       1.090 &       1.078 &       1.158  \\ 
$\rm ^{14}C $ &  1.10(-07)  &  1.15(-07)  &  1.14(-07) &   1.05(-07) &   1.00(-07) &   9.82(-08) &   9.78(-08)  \\
$\rm ^{22}Na$ &  6.95(-08)  &  6.84(-08)  &  5.94(-08) &   5.35(-08) &   4.98(-08) &   4.54(-08) &   4.36(-08)  \\
$\rm ^{26}Al$ &  1.15(-05)  &  1.15(-05)  &  1.09(-05) &   1.06(-05) &   1.03(-05) &   9.88(-06) &   9.68(-06)  \\
$\rm ^{32}Si$ &  9.94(-08)  &  1.55(-07)  &  2.41(-07) &   2.23(-07) &   1.95(-07) &   1.60(-07) &   1.46(-07)  \\
$\rm ^{36}Cl$ &  1.39(-06)  &  1.38(-06)  &  1.32(-06) &   1.34(-06) &   1.35(-06) &   1.38(-06) &   1.39(-06)  \\
$\rm ^{39}Ar$ &  6.47(-07)  &  6.71(-07)  &  6.76(-07) &   6.49(-07) &   6.27(-07) &   5.98(-07) &   5.85(-07)  \\
\enddata         
\end{deluxetable}

\begin{deluxetable}{lrrrrrr}\label{$\rm PF(^{16}O)$}
\tabletypesize{\scriptsize}
\tablewidth{0pt}
\tablecaption{$\rm PF(^{16}O)$ provided by various authors}
\tablehead{
\colhead{Reference} & \colhead{13} & \colhead{15} & \colhead{20} & \colhead{25} & \colhead{30} & \colhead{35}
}
\startdata
Present &  2.3  &  3.8  &  7.6  &  10.9  & 14.0  & 15.4   \\
RHHW02  &       &  6.6  & 12.4  &  15.0  &       &        \\
TNH96   &  1.8  &  3.2  &  8.3  &  13.3  &       &        \\
WW95    &  2.4  &  5.2  & 11.2  &  14.7  & 14.7  & 11.5   \\
\enddata
\end{deluxetable}

\begin{deluxetable}{lccccccccccccc}
\tabletypesize{\scriptsize}
\tablewidth{18cm}
\tablecaption{Percentage Yields coming from the various burnings}
\tablehead{
\colhead{Isotope} &
\colhead{(Expl)}  &
\colhead{(Hec)}   &
\colhead{(Csh)}   &
\colhead{(Expl)}  &
\colhead{(Hec)}   &
\colhead{(Csh)}   &
\colhead{(Expl)}  &
\colhead{(Hec)}   &
\colhead{(Csh)}   &
\colhead{(Expl)}  &
\colhead{(Hec)}   &
\colhead{(Csh)}   &
\colhead{(Expl)} \\
\colhead{} & \colhead{13-15} & \colhead{} & \colhead{20} & \colhead{} & \colhead{} & \colhead{25} & \colhead{} &
\colhead{} & \colhead{30} & \colhead{} & \colhead{} & \colhead{35} & \colhead{}
} 
\startdata  
 Cu63 &100 &  10 &  90 &     &  30 &  60 &  10 &  50 &  50 &     &  60 &  40 &     \\
 Cu65 &100 &     &  60 &  40 &  70 &     &  30 &     & 100 &     &  80 &  -1 &  20 \\
 Zn64 &100 &  80 &     &  20 &  70 &     &  30 & 100 & -50 &     & 100 & -50 &     \\
 Zn66 &100 &  80 &     &  20 &  70 &     &  30 &     & 100 &     &  90 &  10 &     \\
 Zn67 &100 &     & 100 &     &  40 &  60 &     &  30 &  70 &     &  60 &  40 &     \\
 Zn68 &100 &  10 &  50 &  40 &  30 &  50 &  20 &  50 &  40 &  10 &  10 &  70 &  20 \\
 Zn70 &100 &     &     & 100 &     &  20 &  80 &     &  40 &  60 &     &     & 100 \\
 Ga69 &100 &  10 &  70 &  20 &  40 &  40 &  20 &  50 &  40 &  10 &  70 &  20 &  10 \\
\enddata
\end{deluxetable}

\end{document}